\RequirePackage{fix-cm}
\documentclass[natbib,smallextended]{svjour3}       % onecolumn (second format)
\smartqed  % flush right qed marks, e.g. at end of proof
\usepackage{graphicx}
 \journalname{my journal}
\def\aap{Astron.\  Astrophys.\ }

\def\apj{Astroph.\  J.\ }
\def\apjl{Astroph.\  J.\ Lett.\ }

\def\jcp{J. Chem.\ Phys.\ }

\def\pra{Phys.\ Rev.\ A }
\def\prd{Phys.\ Rev.\ D }
\def\prl{Phys.\ Rev.\ Lett.\ }

\def\jms{J. Mol.\ Spectrosc.\ }

\def\lrr{Living\ Rev.\ Relativity }

\def\mnras{Mon.\ Not.\ Roy.\ Astron.\ Soc.\ }

\begin{document}

\title{Search for varying constants of nature from astronomical observation of molecules}

\titlerunning{Search for varying constants of nature}        % if too long for running head

\author{Wim Ubachs
}


\institute{W. Ubachs \at
              Department of Physics and Astronomy, \\
              Vrije Universiteit Amsterdam, \\
              De Boelelaan 1081, \\
              1081 HV Amsterdam, The Netherlands \\
              Tel.: +31-20-5987948\\
              \email{w.m.g.ubachs@vu.nl}
}

\date{Received: date / Accepted: date}

\maketitle

\begin{abstract}
The status of searches for possible variation in the constants of nature from astronomical observation of molecules is reviewed, focusing on the dimensionless constant representing the proton-electron mass ratio $\mu=m_p/m_e$. The optical detection of H$_2$ and CO molecules with large ground-based telescopes (as the ESO-VLT and the Keck telescopes), as well as the detection of H$_2$ with the Cosmic Origins Spectrograph aboard the Hubble Space Telescope is discussed in the context of varying constants, and in connection to different theoretical scenarios. Radio astronomy provides an alternative search strategy bearing the advantage that molecules as NH$_3$ (ammonia) and CH$_3$OH (methanol) can be used, which are much more sensitive to a varying $\mu$ than diatomic molecules. Current constraints are $|\Delta\mu/\mu| < 5 \times 10^{-6}$ for redshift $z=2.0-4.2$, corresponding to look-back times of 10-12.5 Gyrs, and $|\Delta\mu/\mu| < 1.5 \times 10^{-7}$ for $z=0.88$, corresponding to half the age of the Universe (both at 3$\sigma$ statistical significance). Existing bottlenecks and prospects for future improvement with novel instrumentation are discussed.

\end{abstract}

\section{Introduction}
\label{intro}

Extensions of the Standard Model (SM) of physics can occur in various ways. One possibility of such an extension is the temporal or spatial variation of fundamental constants of nature, such as the coupling strengths of the known four forces, the strong and weak forces, electromagnetism and gravity. Their coupling strengths are inserted in the SM as free parameters, as are the relative masses of the elementary particles. However, in order to describe physics in a consistent framework of quantum field theories it does not suffice to allow the fundamental constants to vary at will. For consistency, and to retain e.g. conservation of energy, the inclusion of additional fields is required, often referred to as 'dilaton' fields that may in the simplest form be of scalar nature~\citep{Barrow2002a}. Such fields may be regarded as additional forces or 'quintessence', and may be associated with other unknown forms of (dark) matter or energy. If such connection exists, then the values of some constants of nature might follow the increasing amount of dark energy in the Universe, in which case the variation of constants would manifest itself on a cosmological time scale. Alternatively, the fundamental constants might be connected to local effects such as matter density~\citep{Khoury2004} or gravitational fields~\citep{Flambaum2008}.
Theories involving variation of fundamental constants exist in various flavors and are e.g. described in the review by~\citet{Uzan2011}.

The quantum mechanical energy level structure of atoms and molecules is determined by the value of the fine structure constant $\alpha$ and the proton-to-electron mass ratio $\mu=m_p/m_e$~\citep{Kozlov2013b,Jansen2014}. Searches for variation of both dimensionless constants $\alpha$ and $\mu$ focus primarily on spectroscopic methods. Comparison between spectra obtained from astronomical observations at high redshift with the same spectra measured at high precision in the present epoch with laboratory-based methods may reveal minute shifts in the values of those constants. Spectroscopy, i.e. the measurement of wavelengths or frequencies of absorption lines, is a solid method in view of the accuracy at which such quantities can be measured. In the analysis procedures of comparing spectra the small differential shifts of absorption lines imposed by varying constants must be disentangled from the cosmological and gravitational redshifts involved in astronomical observations. However, General Relativity (GR) predicts that these latter redshifts are dispersionless, hence disentangling is straightforward. As variation of constants implies a breakdown of Einstein's equivalence principle, a corner stone of existing theories of physics, any resulting differential or dispersion effect may be considered as a token for new physics.

The search for a varying fine structure constant $\alpha$ from comparison of spectroscopic lines was pursued already in the 1950s~\citep{Savedoff1956} based on the alkali-doublet method. Here it is used that the separation of two spectral lines probing a spin-orbit doublet from a common ground state scales with $\alpha^2$. \citet{Bahcall1967} therewith constrained a relative variation of $\alpha$ at the $10^{-3}$ level from observations of a pair O III emission lines at redshift $z=0.2$.
The accuracy was much improved by the Many-Multiplet-(MM) method, where absolute transition frequencies of various elements can be used simultaneously for probing a variation of $\alpha$. The MM-method requires detailed ab initio calculations of atomic structure to derive sensitivity coefficients~\citep{Dzuba1999}. Such sensitivity coefficients reflect by how much spectral lines will shift as a result of a relative change in the fine structure constant $\Delta\alpha/\alpha$, and they appear to be largest for heavy elements where electrons undergo relativistic motion. Based on the MM-method and observations with large 8-10m class telescopes  certain indications of a temporally~\citep{Webb2001} and spatially~\citep{Webb2011} varying $\alpha$ were reported. It is discussed in the scientific literature whether these effects might be due, at least in part, to calibration issues with the spectrographs connected to the large optical telescopes used to probe varying $\alpha$ at high redshift~\citep{Rahmani2013,Whitmore2015,Dumont2017}.

Focus of this paper is on another dimensionless constant, the proton-to-electron mass ratio $\mu$, and the constraints that can be derived on its relative variation $\Delta\mu/\mu$. In theoretical models based on Grand Unification it was shown that the rate of change in $\mu$ is much faster than a putative rate of change in $\alpha$~\citep{Calmet2002,Langacker2002}, thus making $\mu$ a sensitive test ground for probing varying constants. Typically the constant $\mu$ is associated with the quantized motion of nuclei in molecules, through their intramolecular rotation, vibration and tunneling modes. Again for any model comparing astronomical spectra with zero-redshift laboratory data, sensitivity coefficients $K_i = (d \ln \lambda)/(d \ln \mu$) must be evaluated for each allowed quantum transition in a molecule~\citep{Jansen2014}. In the following the results of ground-based observation of molecular hydrogen and carbon monoxide at high-redshift will be presented, yielding constraints on $\Delta\mu/\mu$. It will further be discussed how vacuum-ultraviolet (VUV) absorption by H$_2$ in galactic white dwarf objects, detected by a satellite-based spectrograph, will be used for constraining a possible dependence of $\mu$ on a gravitational field. The microwave detection of molecular absorption used for probing a varying $\mu$ is described in a subsequent section. Future developments in optical and radio astronomy, as well as bottlenecks, challenges and opportunities for more tight constraints on variation of the proton-electron mass ratio is discussed in a concluding section.

\section{Optical detection of the H$_2$ molecule}
\label{H2optic}

\begin{table}
\caption{List of the ten high redshift H$_2$ absorption systems analyzed so far for $\mu$-variation. The identification of the quasar source, the redshift $z$ of the absorbing system, and the J2000-coordinates of the quasar source are listed. The column densities $N$(H$_2$) are given on a $\log_{\rm{10}}$ scale in cm$^{-2}$. $R_{mag}$ represents the magnitude in the visible range.}
\label{Table-quasars} %\\
\begin{tabular}{p{2.7cm}p{1.1cm}p{2.0cm}p{2.2cm}p{1.4cm}p{1.1cm}p{1.9cm}}
\hline \hline
Quasar & $z_{\rm{abs}}$ &  RA(J2000)  & Decl.(J2000) & $N$(H$_2$)   & $R_{\rm{mag}}$ & Refs. \\ \hline
\textbf{HE0027$-$1836}& $2.42$  &  00:30:23.62 & $-$18:19:56.0 & $17.3$  & $17.37$  &   \citep{Rahmani2013}  \\
\textbf{Q0347$-$383} & $3.02$  & 03:49:43.64 & $-$38:10:30.6 & $14.5$  &$17.48$   & \citep{Ivanchik2005,Reinhold2006,King2008,Wendt2012,Thompson2009} \\
\textbf{Q0405$-$443} & $2.59$   & 04:07:18.08 & $-$44:10:13.9 & $18.2$  & $17.34$  &    \citep{Ivanchik2005,Reinhold2006,King2008,Thompson2009}   \\
\textbf{Q0528$-$250} & $2.81$   & 05:30:07.95 & $-$25:03:29.7 & $18.2$  & $17.37$  &  \citep{King2008,King2011}  \\
\textbf{B0642$-$5038} & $2.66$   & 06:43:26.99 & $-$50:41:12.7 & $18.4$  & $18.06$  &   \citep{Bagdonaite2014a,Albornoz2014} \\
\textbf{Q1232$+$082} & $2.34$   & 12:34:37.58 & $+$07:58:43.6  & $19.7$  & $18.40$  & \citep{Varshalovich2001,Ivanchik2010,Dapra2017}\\
\textbf{J1237$+$064} & $2.69$   & 12:37:14.60 & $+$06:47:59.5 & $19.2$ &  $18.21$ &
\citep{Dapra2015} \\
\textbf{J1443$+$2724} & $4.22$   & 14:43:31.18 & $+$27:24:36.4 & $18.3$  & $18.81$  &
\citep{Bagdonaite2015}  \\
\textbf{J2123$-$0050} & $2.06$   & 21:23:29.46 & $-$00:50:52.9 & $17.6$  & $15.83$  &
 \citep{Malec2010,Weerdenburg2011} \\
\textbf{Q2348$-$011} & $2.42$   & 23:50:57.87 &  $-$00:52:09.9 & $18.4$   & $18.31$  &
\citep{Bagdonaite2012}\\
\hline
\end{tabular}
\end{table}

Hydrogen is by far the most abundant species in Galaxies and is found in both atomic and molecular forms in the interstellar medium. The investigation of H$_2$ absorption in galaxies at high redshift in the line-of-sight towards quasars favorably draws from this omnipresence of H$_2$. Nevertheless, although large numbers of quasars have been detected, many of which with a Damped-Lyman-$\alpha$ (DLA) absorption feature marking a column density for atomic hydrogen in excess of $2 \times 10^{20}$ cm$^{-2}$, only a limited number of systems with pronounced H$_2$ absorption have been identified. Currently some 23 H$_2$ absorbers at redshifts $z>2$ have been found by various authors and a compilation can be found in~\citet{Ubachs2016}. An additional set of 23 systems with tentative absorptions have been detected~\citep{Balashev2014}. In those systems typically a column density for H$_2$ is detected at the 1\% level with respect to that of H, although much higher molecular fractions have also been detected recently, see e.g. \cite{Noterdaeme2017}.

The strongest electric dipole-allowed transitions in the H$_2$ molecule, those of the Lyman and Werner band systems, occur at rest-frame wavelengths of 912-1150 \AA, so that a significant fraction of the absorption lines shift into the atmospheric transmission window ($\lambda > 3050$ \AA) for absorption redshifts of $z>2$. Since the molecular absorption bands lay shortward in wavelength of the Lyman-$\alpha$ line ($1215$ \AA), H$_2$ line absorptions are typically found in the 'Lyman-$\alpha$ forest'. This consists of a spectral region crowded with many Lyman-$\alpha$ absorption lines, produced by the intergalactic gas over the full range of redshifts along the line of sight, from the observer to the background quasar \citep{Rauch1998}.
This Lyman-$\alpha$ forest produces in fact a random spectrum overlaying the spectral region where H$_2$ absorption lines are located, thus forming an obstacle to be overcome in the data analysis~\citep{Malec2010}.

Currently some ten H$_2$ absorptions systems, listed in Table~\ref{Table-quasars}, have been analyzed for $\mu$-variation. These are good quality systems in view of brightness of the background quasar, typically in the range of $R_{mag}= 17.3 - 19$. Such magnitude allows for recording of an H$_2$ absorption spectrum of typical SNR $\geq 20$ and resolution of $R=50,000$ in an amount of 10 hrs at an 8-10m class telescope, such as the Keck telescope (with the HIRES spectrograph) and the Very Large Telescope (VLT, with the UVES spectrograph). The quasar J2123-005 is a special case displaying extraordinary high brightness ($R_{mag}=15.83$), which was exploited in observations by both telescopes~\citep{Malec2010,Weerdenburg2011}. Further, good quality H$_2$ absorption spectra can be obtained for column densities of $\log_{\rm{10}}N$(H$_2$) [cm$^{-2}$] $= 17-18.5$. System Q0347-383 with $\log_{\rm{10}}N$(H$_2$) [cm$^{-2}$] $= 14.5$  produces nevertheless a good quality spectrum with non-saturated lines \citep{Ivanchik2005,Wendt2012}, while system J1237+064 exhibits strongly saturated absorption lines for the highest populated quantum states of H$_2$~\citep{Noterdaeme2010}.

\begin{figure}%[!ht]
\centering
\includegraphics[scale=0.7]{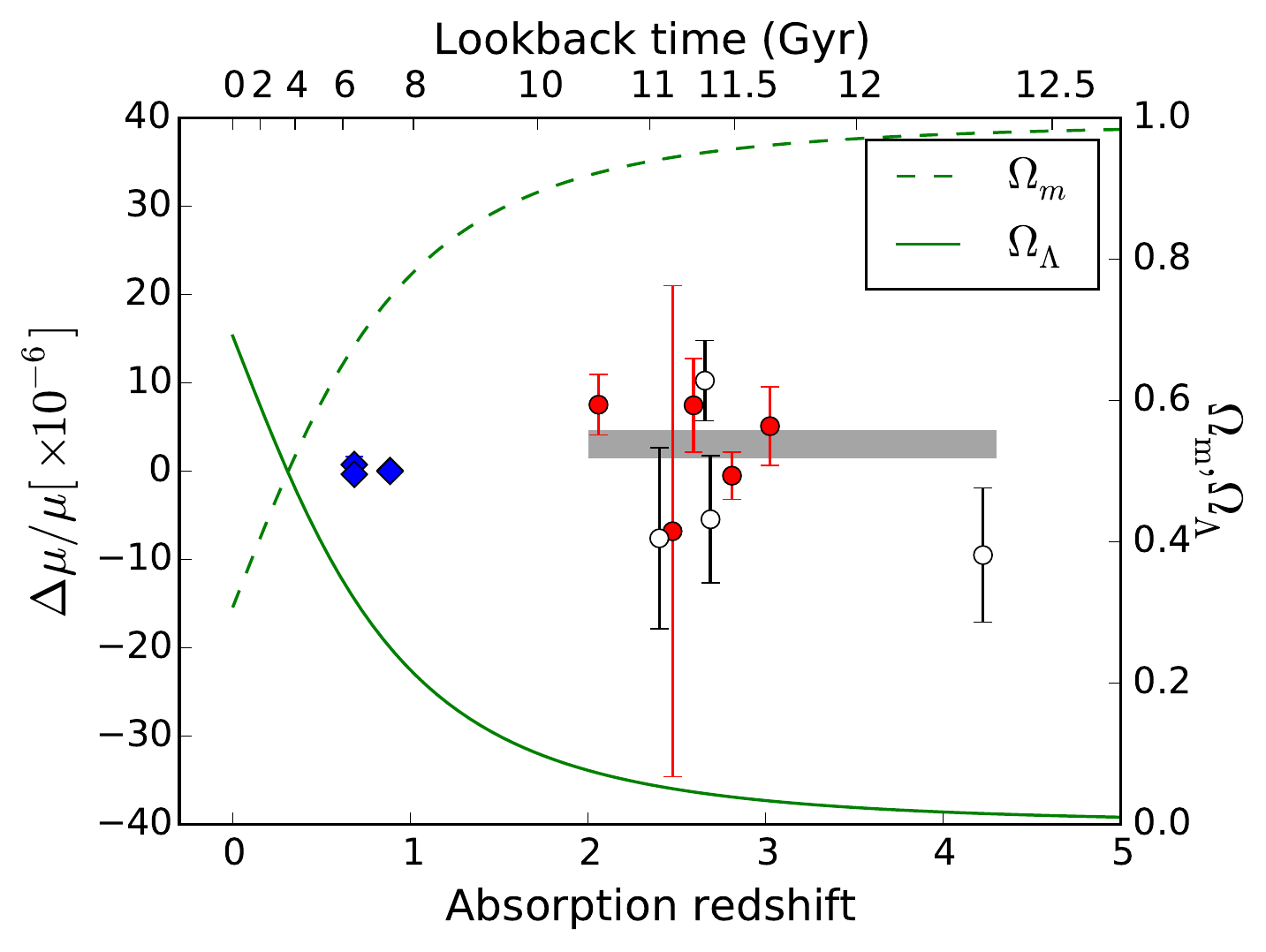}
\caption{Values for the relative variation of the proton-to-electron mass ratio $\Delta\mu/\mu$ plotted as a function of redshift and look-back time for nine H$_2$ absorption systems analyzed in detail (Q1232+082 not included). Open circles: data corrected for long-range wavelength distortions; Closed (red) circles: uncorrected data. Results are plotted alongside with evolution of cosmological parameters for $\Omega_m$ (dashed line) and $\Omega_{\Lambda}$ (full line), with respect to the right-hand vertical axis. The grey bar represents the average for $\Delta\mu/\mu$ with $\pm 1\sigma$ uncertainty limits from the H$_2$ systems analyzed. For comparison results from radio astronomical observations at lower redshift ($z<1$) are plotted as well (diamonds - blue). Figure reproduced from~\citet{Ubachs2016}.}
\label{mu-plot-overview}
\end{figure}

Quasar absorption spectra are treated in a 'comprehensive fitting' analysis~\citep{Malec2010}, where all H$_2$ lines are described by a set of molecular parameters, $\lambda_i^{\rm{0}}$ for the rest-frame wavelength, $f_i$ for the line oscillator strength, and $\Gamma_i$ for the natural broadening coefficient, and the sensitivity coefficients $K_i$ and invoked in the equation governing the comparison between spectra:
\begin{equation}
\lambda^z_i= \lambda^{\rm{0}}_i (1+z_{\rm{abs}})(1+K_i\frac{\Delta\mu}{\mu}),
\label{vpfit-equation}
\end{equation}
The rest frame wavelengths $\lambda_i^{\rm{0}}$ of the H$_2$ absorption lines have been determined to high accuracy in dedicated laboratory experiments using tunable vacuum ultraviolet lasers \citep{Ubachs2004,Philip2004}, as well as by a combination of spectroscopic studies aimed at determining level energies of excited states in H$_2$ from which the vacuum ultraviolet absorption lines can be calculated at high accuracy \citep{Bailly2010}. Also lines of the HD isotopologue are used in the astronomical analyses for which accurate laboratory data are available \citep{Ivanov2008a}. The status of these laboratory experiments is that the transition wavelengths of the H$_2$ absorptions may be considered exact for the purpose of comparison to quasar data recorded with the HIRES spectrometer at Keck and the UVES spectrometer at VLT. The values $f_i$ and $\Gamma_i$ for the absorption  lines are obtained from ab initio calculations of the H$_2$ molecule \citep{Abgrall2000}. Values for the sensitivity coefficients $K_i$ were derived by a number of groups using independent techniques \citep{Varshalovich1993,Meshkov2006,Ubachs2007}.

In the analysis of astronomical spectra each spectral line is folded by a Doppler parameter $b$, shifted by the redshift parameter $1+z$ and its intensity multiplied by the column density $N_J$ for each populated quantum state of the H$_2$ molecule. Typically only states with $v=0$ and $J=0-5$ will be populated to the extent that absorption is detected. Hence, from the broadening parameters $\Gamma_i$ and $b$, further folded by the instrument function of the spectrograph, a Voigt profile shape for the lines is obtained. While applying the fitting routine to the entire spectrum multiple velocity components, introduced as multiple redshift paramaters $z_i$, will be allowed for optimizing the entire spectrum. In this analysis the occurrence of Lyman-$\alpha$ forest lines as well as metal line absorption will be included. In the final stage of this comprehensive fitting method, after a well-behaved absorption model is created, $\Delta\mu/\mu$ is introduced as a final free parameter and fitted.

This procedure has been applied to several of the ten absorption systems listed in Table~\ref{Table-quasars}, while for some other systems a 'line-by-line' fitting model was used~\citep{Thompson2009,Wendt2012}. The latter method exhibits some drawbacks when dealing with overlapped lines in the fit~\citep{Malec2010,King2011}. Results for the values $\Delta\mu/\mu$ obtained from the various systems at redshifts in the range $z=2.0-4.2$ analyzed have been collected in Fig.~\ref{mu-plot-overview}. A weighted average yields $\Delta\mu/\mu = (3.1 \pm 1.6)\times 10^{-6}$. This result, indicating a small 1.9$\sigma$ effect of a varying $\mu$ is, in view of its small significance, treated as a null-result. This null-result may also be phrased as $|\Delta\mu/\mu| <  5 \times 10^{-6}$ at a 3$\sigma$ significance level~\citep{Ubachs2016}.

\section{Optical detection of the CO molecule}
\label{COoptic}

Carbon monoxide is the second most abundant gaseous molecule in the Universe and its A-X electronic band system, occurring at rest-frame wavelengths of $\lambda < 1544$ \AA, is observed in a number of quasar absorption systems~\citep{Noterdaeme2008,Noterdaeme2010,Noterdaeme2011,Noterdaeme2017}. Since the wavelength interval of the A-X bands is longward of the Lyman-$\alpha$ line these CO spectra are not overlaid by Lyman-$\alpha$ forest lines.

\begin{figure}%[!ht]
\centering
\includegraphics[scale=0.4]{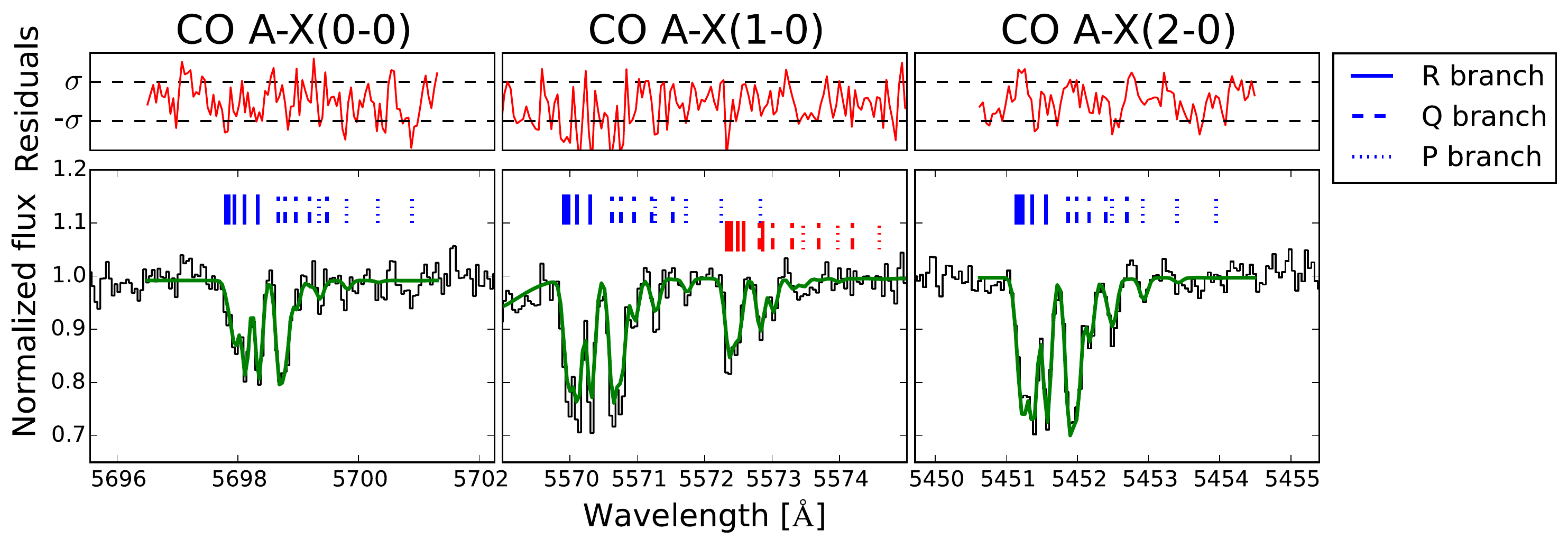}
\caption{Spectra of the A-X ($v$,0) bands, for $v=0-2$, observed in the line of sight toward J1237+064 at redshift $z=2.69$. In the upper panels the residuals from a fit are displayed.}
\label{COspec}
\end{figure}

In the quasar absorption system J1237+064  CO is detected at $z=2.69$ (with ESO-VLT) with a column density of $\log_{\rm{10}}N$(CO)[cm$^{-2}$] $= 14.2$ and absorption bands A-X($v$,0) for $v=0-8$ were observed in the range $\lambda= 5696-4880$ \AA. In addition also the B-X(0,0), C-X(0,0) and E-X(0,0) bands were observed in the bluer part of the spectrum. Molecular parameters were measured in laser-based and synchrotron-based spectroscopic absorption studies~\citep{Salumbides2012} and invoked in the 'comprehensive fitting' method based on Eq.~(\ref{vpfit-equation}). In Fig.~\ref{COspec} spectra of the three lowest bands of CO A-X, as observed in J1237+064, are displayed with a model fit. The analysis of the CO spectrum yields a value of $\Delta\mu/\mu = (2.3 \pm 1.7)\times 10^{-5}$~\citep{Dapra2016}. This shows that the CO method is somewhat less constraining than the H$_2$ method, which is ascribed to the fact that the details of the CO absorption features are spread over less pixels in the spectrum than those of H$_2$. The range of $K$-coefficients is similar for both molecules. Recently also $\mu$-variation was analyzed from a second absorber, J0000+0048 at $z=2.52$, resulting in $\Delta\mu/\mu = (1.8 \pm 2.3)\times 10^{-5}$~\citep{Dapra2017a}.

\section{Detection of H$_2$ in the photosphere of white dwarfs}
\label{vuv-method}

The spectrum of molecular hydrogen can also be used to search for a possible dependence of the fundamental constant $\mu$ on a gravitational field.  Theoretically such phenomenon might be explained in 'chameleon' scenarios of varying constants~\citep{Khoury2004,Flambaum2008}. Some white dwarfs (WD) exhibit a thin photosphere layer in which molecular hydrogen molecules give rise to an absorption spectrum~\citep{Xu2013}. The conditions applying to the H$_2$ molecules in such circumstances are extreme. Since white dwarfs are collapsed stars of typically a solar mass shrunk to a radius $R$ of 1\% of the Sun, the gravitational potential $\phi=GM/Rc^2$ is about $10\,000$ times that on the Earth's surface. Further, the temperature of the photosphere is typically in excess of $10\,000$ K. Spectra of the photosphere of two white dwarfs, GD-29-38 (or WD2326+049)  and GD-133  (WD1116+026) were observed with the Cosmic  Origins Spectrograph (COS) aboard the Hubble Space Telescope (HST).

\begin{figure}%[!ht]
\centering
\includegraphics[scale=0.5]{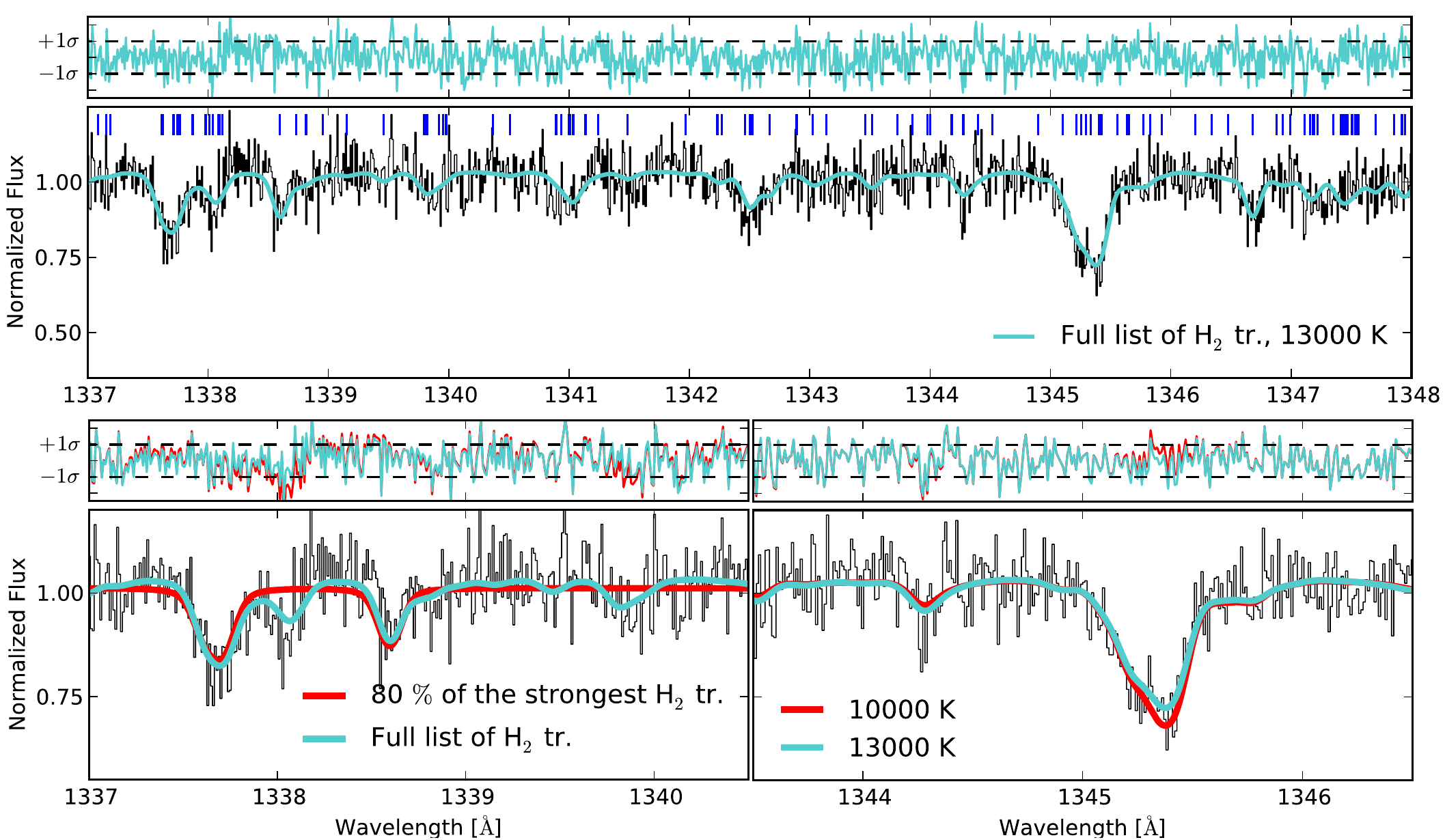}
\caption{Part of the absorption spectrum of the photosphere of the white dwarf system GD29-38, overlaid with a model spectrum for temperatures of $T=10\,000$ K and using the 80\% strongest lines (red line), and a model for $T=13\,000$ K and the full list of H$_2$ absorption lines (blue-green line). The total spectrum used in the analysis extends from $1310 - 1412$ \AA. }
\label{WDspec}
\end{figure}

Studies aiming to investigate gravitational effects in dim objects as WD stars can preferably be performed on systems observed inside the Galaxy. However, in such case there will be no redshift to shift the absorption profiles into the atmospheric transmission window. Galactic WD spectra should then be recorded in the vacuum ultraviolet range, for which COS-HST is ideally suited. An example of a Galactic WD H$_2$ absorption spectrum is displayed in Fig.~\ref{WDspec} for the case of GD-29-38. In view of the high local temperatures higher vibrational states of H$_2$ are populated in the WD photospheres, in contrast to the cold environments probed in quasar absorption, where only H$_2$($v=0$) is detected. In view of the Franck-Condon factors of H$_2$ the Lyman bands B-X(1,3) and (1,4) are the most strongly represented in the observed spectrum which covers the range $1310-1412$ \AA~\citep{Bagdonaite2014b}. At temperatures of $T>10\,000$ K rotational quantum states of up to $J=25$ are effectively populated and a simulation of the spectra involves starting from the partition function including all states populated~\citep{Salumbides2015}:
\begin{equation}
P_{v,J}(T) = \frac{g_n (2J+1) e^{\frac{-E_{v,J}}{kT}}}{\sum\limits_{v=0}^{v_{\mathrm{max}}} \sum\limits_{J=0}^{J_{\mathrm{max}}(v)} g_n (2J+1) e^{\frac{-E_{v,J}}{kT}}}
\label{partition-function}
\end{equation}
Based on this partition function and assuming thermodynamic equilibrium for the H$_2$ populated quantum states a fit can be made to the observed spectra, with temperature $T$ and $\Delta\mu/\mu$ as free parameters, again using Eq.~(\ref{vpfit-equation}) for the absorption model. This method results in constraints on dependences of $\mu$ on a gravitational field quantified as $\Delta\mu/\mu = (-5.8 \pm 3.9) \times 10^{-5}$ for GD-29-38 (at $\phi =  1.9 \times 10^{-4}$) and $\Delta\mu/\mu = (-2.7 \pm 4.9) \times 10^{-5}$ for GD-133 (at $\phi =  1.2 \times 10^{-4}$)~\citep{Bagdonaite2014b}. Thus far the objects GD-29-38 and GD-133 are the only white dwarf stars for which an H$_2$ absorption spectrum of sufficient SNR was obtained. In a third system, GD-31~\citep{Xu2013}, H$_2$ was observed but the spectrum is of too low quality to derive a competitive constraint on $\mu$.

\section{Radio astronomical observation of polyatomic molecules}
\label{radio-astronomy}

\begin{figure}%[!ht]
\centering
\includegraphics[scale=0.86]{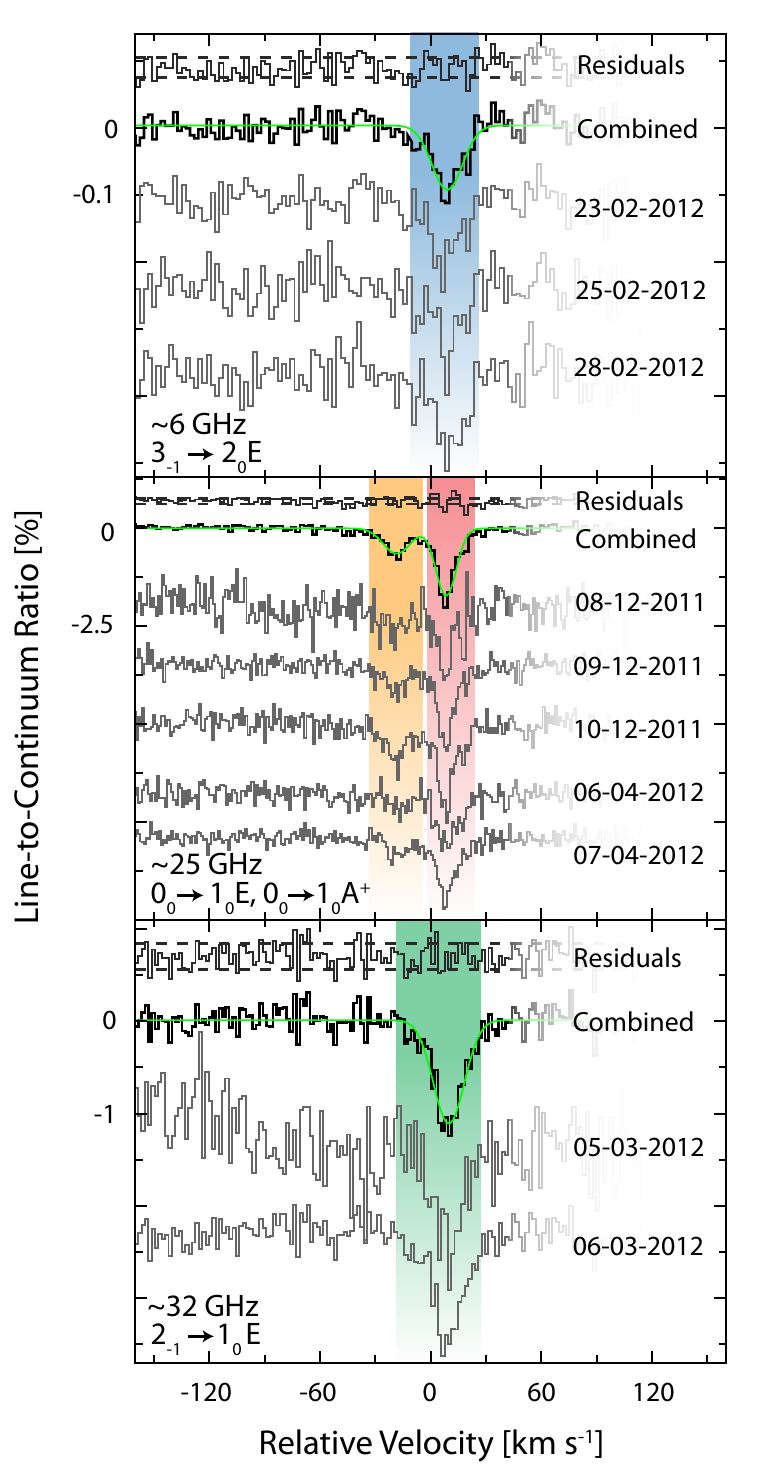}
\caption{The four methanol absorption lines observed with the Effelsberg radio telescope at $z=0.88582$ towards the radio-loud quasar PKS1830-211. Figure adapted from~\citet{Bagdonaite2013a}.}
\label{Meth}
\end{figure}

Radio astronomy is the preferred technique for identifying molecules in the interstellar medium and currently well over a 100 molecular species have been detected through observations in our Galaxy. Only a limited sample of extragalactic radio sources at intermediate redshift ($0.2 < z <1$) with molecular absorption in its sightline have been identified, the four listed in Table~\ref{Radio-sources} which have been used to search for varying constants on a cosmological time scale, and a fifth system B1504$+$377 in which OH has been detected~\citep{Kanekar2002}. The radio-loud quasars B0218+357 and PKS 1830-211  are strongly lensed radio galaxies with the radio source far in the background, exhibit a large extinction, $A_v \sim 10-100$, and thus a very high molecular fraction.

Most of the molecules detected in the radio range are observed through pure rotational lines in the spectrum, for which the sensitivity coefficients are $K_{\mu}=-1$~\citep{Jansen2014}. This implies that all such transitions exhibit the same sensitivity to $\mu$-variation. Hence $\mu$-variation does not produce a differential effect, to be distinguished from a cosmological or a Doppler shift, not even if transitions of different molecules are included in a combined analysis. If however molecules are considered that exhibit other internal modes of energy a favorable situation arises. The ammonia molecule exhibits an inversion splitting associated with a tunneling inversion mode (the N-atom tunneling through the plane of the three H-atoms) which is very sensitive to a mass variation, in view of quantum tunneling being exponentially dependent on mass. This gives rise to a sensitivity coefficient of $K_{\mu}=-4.2$ for the typical 21 GHz lines of NH$_3$~\citep{Flambaum2007}. However, again the various rotational-tunneling lines all exhibit the same $K$-value, so that purely rotational lines of other molecules (at $K_{\mu}=-1$) must be included to derive constraints on $\Delta\mu/\mu$. Typically HC$_3$N lines can be used for that purpose~\citep{Henkel2009}, but the analysis then is based on the assumption that NH$_3$ and HC$_3$N  molecules reside in the same location in the extragalactic absorbing clouds. Ammonia has thusfar been observed in two intermediate redshift objects, PKS1830-211~\citep{Henkel2009} and B0218+357~\citep{Murphy2008a}.

The methanol molecule is an exemplary case where these shortcomings are circumvented. The motion of hindered internal rotation of the OH-group over the CH$_3$ group gives rise to a tunneling motion for which some rotational-torsional transitions exhibit very large $K_{\mu}$ coefficients, with values of $K_{\mu}=-7.4$ and $K_{\mu}=-33$~\citep{Jansen2011,Levshakov2011}. When these transitions are combined with pure rotational transitions, of the same molecular species, then a very sensitive analysis of $\mu$-variation can be performed. This was done in an observation of four radio lines in the methanol spectrum, ranging from 6.5 GHz to 32 GHz, with the Effelsberg 100m dish radio telescope (see Fig.~\ref{Meth}~\citep{Bagdonaite2013a}). After that initial study more extended observations of methanol in the same object have been observed, using also the IRAM30m telescope and the ALMA telescope array~\citep{Bagdonaite2013b}, as well as the Extended Very Large Array~\citep{Kanekar2015}. These studies have produced the tightest constraint of $|\Delta\mu/\mu| < 10^{-7}$, however only for a single object and at a single redshift, equivalent to 7.5 Gyrs look-back time or half the age of the Universe. The stringent constraints on $\Delta\mu/\mu$ derived from NH$_3$ and CH$_3$OH at redshifts $z=0.68$ and $z=0.88$ are included in Fig.~\ref{mu-plot-overview}.

\begin{table}
\caption{List of  extra-galactic radio sources with molecules detected in absorption searching for varying constants.}
\label{Radio-sources} %\\
\begin{tabular}{p{3.5cm}p{1.4cm}p{2.0cm}p{2.2cm}p{3.4cm}}
\hline \hline
Radio Quasar &  $z_{\rm{abs}}$   &  RA(J2000)  & Decl.(J2000)      & Refs. \\ \hline
\textbf{PKS1830$-$211}  & $0.882$   & 18:33:39.9  & $-$21:03:40    & \citep{Henkel2009,Bagdonaite2013a,Bagdonaite2013b,Kanekar2015}    \\
\textbf{B0218$+$357}    & $0.684$   & 02:21:05.5  & $+$35:56:14    & \citep{Murphy2008a,Kanekar2011}    \\
\textbf{PKS1413$+$135}  & $0.247$   & 14:15:58.8  & $+$13:20:24    & \citep{Darling2004}   \\
%\textbf{B1504$+$377}    & $0.671$   & 15:06:09.5  & $+$37:30:51    &    \\
\textbf{PMN0134$-$0931} & $0.765$   & 01:34:35.7  & $-$09:31:03    & \citep{Kanekar2005}   \\
\hline
\end{tabular}
\end{table}

In two other radio systems PMN J0134-0931 with a lensed galaxy at $z = 0.765$, and PKS1413+135 absorption of OH radicals was analyzed that provided combined constraints on $\mu$ and $\alpha$. The observation of OH lines at 18 cm was combined with that of the HI 21 cm line, from which contributions to varying $\mu$ and $\alpha$ may be disentangled~\citep{Darling2004,Kanekar2005}.

Most investigations on varying constants focus on absorption, because cold environments may be probed without too much turbulent motion and line broadening. It was shown however, that emission of CO combined with emission of atomic C I, towards the lensed galaxy  HLSJ091828.6+514223 at $z=5.2$  could be employed to derive a constraint on $\mu$-variation at such high redshifts~\citep{Levshakov2012}.

\section{Future opportunities and instrumentation}

The accuracy of the H$_2$ and CO optical absorption methods as discussed in sections~\ref{H2optic} and \ref{COoptic} is limited by the signal-to-noise ratio (SNR) obtained in the spectra recorded at the 8-10m class telescopes. Extending the integration time will improve the spectra and therewith the statistical constraint on $\Delta\mu/\mu$, but the SNR scales only with the root of measurement time. Since for the measurement of the H$_2$ absorption spectrum of Q2348-011 already an averaging time of 19 hrs on target was applied~\citep{Bagdonaite2012}, it seems unrealistic to further extend such long observation periods on 8-10m class telescopes. However, the construction of a next generation optical telescopes with 30-40m dishes, such as the European Extremely Large Telescope, the Thirty Meter Telescope and the Giant Magellan Telescope, will deliver the prospect of increased photon collection and drastically improved signal-to-noise ratio. This will straightforwardly lead to tighter bounds on $\Delta\mu/\mu$ with the H$_2$ and CO optical methods.

The spectrographs used for $\mu$-variation studies (UVES at VLT, and HIRES at Keck) are designed such that a resolving power of $R \sim 50\,000$ is obtained for an entrance slit corresponding to 0.8 arcseconds; this value matches 'regular-to-good' seeing conditions so that all light can be captured within the slit at seeing conditions of 0.8". For the molecular hydrogen absorbing systems investigated so far it was found that the Doppler broadening of the individual H$_2$ features in all of the cases yielded superthermal line broadenings in excess of the $50\,000$ resolving power limit, most likely due to turbulent motion inside the absorbing galaxy or unresolved velocity substructures. In the specific case of J2123-005 spectra were measured at $R=55\,000$ with UVES-VLT~\citep{Weerdenburg2011} and $R=110\,000$ with HIRES-Keck~\citep{Malec2010}, but the linewidths in the observed spectra were the same, hence fully governed by superthermal Doppler broadening. In such cases, for all systems investigated (see Table~\ref{Table-quasars}), the effective Doppler width greatly exceeds the kinematic width that would correspond to the Boltzmann population temperatures, which are for H$_2$ typically $50-150$ K~\citep{Petitjean2002,Srianand2005}. Although it might be possible that an H$_2$ absorbing system will be found, exhibiting no turbulent motions and a strongly decreased Doppler broadening, as of now it appears that improved resolution beyond $50\,000$ would probably not benefit the H$_2$ method. The planned installation of the ESPRESSO spectrograph at ESO-VLT with its high-resolution mode of $R=120\,000$ and its very-high-resolution mode of $R=220\,000$ would provide an opportunity to further test this hypothesis.
%The same holds for the high resolution spectrograph on the LUVOIR surveyor \citep{France2016}.

Wavelength calibration is a crucial issue in a spectroscopic search for varying constants, because the methods rely on exploring the smallest differential wavelength shifts between molecular absorption lines. The spectra are generally calibrated against a ThAr reference spectrum measured immediately after the quasar exposures without resetting its geometry. It was found that "intra-order" distortions occur in the echelle-grating based measurements~\citep{Griest2010,Whitmore2010}. In addition, it was found that long-range wavelength distortions also affect the calibration of the quasar absorption spectra when calibrated against the ThAr lines. This is probably due to the fact that the light beams as originating from far-distant objects and the light beams from the ThAr cell on the telescope platform travel along different paths ~\citep{Rahmani2013,Whitmore2015}. The resulting distortions of the wavelength scale may mimic a $\Delta\mu/\mu$ effect. For analyzing these miscalibration phenomena exposures of either solar twin stars or of asteroids are used, which may be compared with an accurate high-resolution Fourier-transform spectra measured of the Sun~\citep{Molaro2008}. Such recalibration methods may be applied to assess a value for the systematic uncertainty in the determination of $\Delta\mu/\mu$, or even counter-distort the wavelength scale to obtain a fiducial value of a constraint on $\Delta\mu/\mu$, as was pursued in some studies~\citep{Rahmani2013,Bagdonaite2014a,Dapra2015}.

Some of the problems of wavelength calibration of astronomical spectra can be resolved by fiber-feeding the light into the spectrograph. Already for the ESPRESSO spectrograph the light from the VLT large-dish will be collected into a multi-mode fiber before entering the spectrograph, with the goal to discriminate against angular differences between light paths. A problem with fiber-feeding is the additional loss of photons, in particular in the UV part of the spectrum. In fact the fiber-fed spectrographs will not be specified for the range $3000-3800$ \AA, a range that contains the central information for the H$_2$ method in case of quasars at redshifts $z=2-3$. Hence fiber feeding the spectrograph will be detrimental in application to most of the H$_2$ absorbers that fall in that redshift range (see Table~\ref{Table-quasars}).

Another improvement, currently established on some telescope-based spectrographs, is the use of frequency comb lasers at the observatories~\citep{Wilken2012}. Frequency comb lasers deliver a very dense spectrum of frequency markers at a separation exactly governed by an atomic clock, thus constituting an ideal reference spectrum. However, extension of the comb markers to the UV part of the spectrum at $3000-3800$ \AA\ is not obvious. Again, a combination with fiber-feeding is anticipated, resulting in similar problems for the H$_2$ method as mentioned above.

The method to detect a dependence of $\mu$ on gravitational fields, as discussed in section~\ref{vuv-method}, requires a high resolution satellite-based VUV spectrograph, since zero-redshift absorption by H$_2$ occurs in the wavelength range for which the Earth's atmosphere is opaque. Currently the COS-spectrograph aboard the HST serves this purpose although for the study of H$_2$ in the Milky Way it would be desirable to use a spectrometer of a resolving power beyond that of COS ($R=20\,000$). Repair on HST instruments is likely to stop after the launch of the James Webb Space Telescope in the near future. The latter takes advantage of positioning in a Lagrange point of the Earth orbit around the sun at 1.5 million km from Earth, but is only equipped with instruments for the infrared. While in the past Galactic molecular hydrogen was first detected from a rocket-sounding mission~\citep{Carruthers1970}, thereafter dedicated VUV-spectrographs were launched in orbit: the Copernicus mission, the Orpheus telescope aboard a Space Shuttle, the Far Ultraviolet Spectroscopic Explorer (FUSE) mission, and the Space Telescope Imaging Spectrograph (STIS) and the Cosmic Origins Spectrograph (COS) instruments aboard the HST. The possibility of detecting and studying H$_2$ within the Milky Way, for any astrophysical purpose, is strongly endangered after HST and its instruments reach the end of their lifetime. A possible follow-up mission is that of the Large/UV/Optical/Infrared (LUVOIR) surveyor covering the H$_2$ absorption wavelength range, which is however still in a preparation phase~\citep{France2016}.
% and launch a high-resolution ($R=50\,000$) VUV spectrograph in orbit for studying Galactic H$_2$.

As for the radio astronomical observations in search for a varying $\mu$, discussed in section~\ref{radio-astronomy}, it is established that probing the methanol molecule has two distinct advantages: CH$_3$OH exhibits the most sensitive transitions and it exhibits transitions with a wide variety of sensitivities, such that there is no need to compare with other molecular species providing anchor lines. However, thusfar only a single absorbing system with methanol toward a radio-loud quasar is found. The object at $z=0.88$ in the line-of-sight toward PKS1830-211 is a lensed galaxy, while the background source is a blazar and exhibiting a variable luminosity~\citep{Muller2011}, which brings some difficulties when studying it in the context of $\mu$-variation~\citep{Bagdonaite2013b}. In a strategy of tightening the limitations on $\Delta\mu/\mu$ the most sensitive radio observatories could be used for finding additional methanol absorbers, preferably at higher redshifts ($z>1$). For such survey studies the Atacama Large Millimeter Array (ALMA) observatory could be used, targeting high frequency transitions of methanol in the ALMA window ($> 100$ GHz). It should be noted however that the roto-torsional modes in methanol, exhibiting the largest sensitivity coefficients, are at lower frequencies, typically at $<50$ GHz, depending on redshift. While currently the Effelsberg 100m telescope~\citep{Bagdonaite2013a} and the Extended Very Large Array~\citep{Kanekar2015} are suitable observatories for the low-frequency range, in future the Square Kilometer Array (SKA) will be a very sensitive instrument to probe sensitive transitions in methanol for further constraining fundamental constants as the proton-electron mass ratio, or find a positive effect of new physics beyond the prevailing Standard Model.

\begin{acknowledgements}
The author acknowledges financial support from the European Research Council (ERC) under the European Union's Horizon 2020 research and innovation programme (Grant Agreement No. 670168).
\end{acknowledgements}


\begin{thebibliography}{75}
% BibTex style file: aps.bst  (nameyear), 2013-04-23
\ifx \bisbn   \undefined \def \bisbn  #1{ISBN #1}\fi
\ifx \binits  \undefined \def \binits#1{#1} \fi
\ifx \bauthor  \undefined \def \bauthor#1{#1} \fi
\ifx \bjtitle  \undefined \def \bjtitle#1{\textrm{#1}}\fi
\ifx \batitle  \undefined \def \batitle#1{#1} \fi
\ifx \bctitle  \undefined \def \bctitle#1{#1} \fi
\ifx \bvolume  \undefined \def \bvolume#1{\textbf{#1}}\fi
\ifx \byear  \undefined \def \byear#1{#1} \fi
\ifx \bissue  \undefined \def \bissue#1{#1} \fi
\ifx \bfpage  \undefined \def \bfpage#1{#1} \fi
\ifx \blpage  \undefined \def \blpage #1{#1} \fi
\ifx \burl  \undefined \def \burl#1{#1} \fi
\ifx \doiurl  \undefined \def \doiurl#1{#1} \fi
\ifx \betal  \undefined \def \betal{et al.} \fi
\ifx \binstitute  \undefined \def \binstitute#1{#1} \fi
\ifx \beditor  \undefined \def \beditor#1{#1} \fi
\ifx \bpublisher  \undefined \def \bpublisher#1{#1} \fi
\ifx \bbtitle  \undefined \def \bbtitle#1{\textit{#1}} \fi
\ifx \bedition  \undefined \def \bedition#1{#1} \fi
\ifx \bseriesno  \undefined \def \bseriesno#1{#1} \fi
\ifx \blocation  \undefined \def \blocation#1{#1} \fi
\ifx \bsertitle  \undefined \def \bsertitle#1{#1} \fi
\ifx \bsnm \undefined \def \bsnm#1{#1} \fi
\ifx \bsuffix \undefined \def \bsuffix#1{#1} \fi
\ifx \bparticle \undefined \def \bparticle#1{#1} \fi
\ifx \barticle \undefined \def \barticle#1{#1} \fi
\ifx \botherref \undefined \def \botherref #1{#1} \fi
\ifx \url \undefined \def \url#1{#1} \fi
\ifx \bchapter \undefined \def \bchapter#1{#1} \fi
\ifx \bbook \undefined \def \bbook#1{#1} \fi
\ifx \bcomment \undefined \def \bcomment#1{#1} \fi
\ifx \oauthor \undefined \def \oauthor#1{#1} \fi
\ifx \citeauthoryear \undefined \def \citeauthoryear#1{#1} \fi
\ifx \texttildelow  \undefined \def \texttildelow{\symbol{126}} \fi
\def \endbibitem {}
\ifx \bconflocation  \undefined \def \bconflocation#1{#1} \fi

\bibitem[\protect\citeauthoryear{{Abgrall} et~al.}{2000}]{Abgrall2000}
\begin{barticle}
\bauthor{\binits{H.} \bsnm{{Abgrall}}},
\bauthor{\binits{E.} \bsnm{{Roueff}}},
\bauthor{\binits{I.} \bsnm{{Drira}}},
\batitle{{Total transition probability and spontaneous radiative dissociation
  of B, C, B$^{'}$ and D states of molecular hydrogen}}.
\bjtitle{Astron. Astrophys. Suppl. Ser.}
\bvolume{141},
\bfpage{297}--\blpage{300}
(\byear{2000})
\end{barticle}
\endbibitem

\bibitem[\protect\citeauthoryear{{Albornoz V{\'a}squez}
  et~al.}{2014}]{Albornoz2014}
\begin{barticle}
\bauthor{\binits{D.} \bsnm{{Albornoz V{\'a}squez}}},
\bauthor{\binits{H.} \bsnm{{Rahmani}}},
\bauthor{\binits{P.} \bsnm{{Noterdaeme}}},
\bauthor{\binits{P.} \bsnm{{Petitjean}}},
\bauthor{\binits{R.} \bsnm{{Srianand}}},
\bauthor{\binits{C.} \bsnm{{Ledoux}}},
\batitle{{Molecular hydrogen in the $z_{abs} = 2.66$ damped Lyman-{$\alpha$}
  absorber towards QJ0643$-$5041. Physical conditions and limits on the
  cosmological variation of the proton-to-electron mass ratio}}.
\bjtitle{\aap}
\bvolume{562},
\bfpage{88}
(\byear{2014})
\end{barticle}
\endbibitem

\bibitem[\protect\citeauthoryear{{Bagdonaite} et~al.}{2012}]{Bagdonaite2012}
\begin{barticle}
\bauthor{\binits{J.} \bsnm{{Bagdonaite}}},
\bauthor{\binits{M.T.} \bsnm{{Murphy}}},
\bauthor{\binits{L.} \bsnm{{Kaper}}},
\bauthor{\binits{W.} \bsnm{{Ubachs}}},
\batitle{{Constraint on a variation of the proton-to-electron mass ratio from
  H$_{2}$ absorption towards quasar Q2348$-$011}}.
\bjtitle{\mnras}
\bvolume{421},
\bfpage{419}--\blpage{425}
(\byear{2012})
\end{barticle}
\endbibitem

\bibitem[\protect\citeauthoryear{{Bagdonaite} et~al.}{2013a}]{Bagdonaite2013a}
\begin{barticle}
\bauthor{\binits{J.} \bsnm{{Bagdonaite}}},
\bauthor{\binits{P.} \bsnm{{Jansen}}},
\bauthor{\binits{C.} \bsnm{{Henkel}}},
\bauthor{\binits{H.L.} \bsnm{{Bethlem}}},
\bauthor{\binits{K.M.} \bsnm{{Menten}}},
\bauthor{\binits{W.} \bsnm{{Ubachs}}},
\batitle{{A stringent limit on a drifting proton-to-electron mass ratio from
  alcohol in the early {U}niverse}}.
\bjtitle{Science}
\bvolume{339},
\bfpage{46}--\blpage{48}
(\byear{2013}a)
\end{barticle}
\endbibitem

\bibitem[\protect\citeauthoryear{{Bagdonaite} et~al.}{2013b}]{Bagdonaite2013b}
\begin{barticle}
\bauthor{\binits{J.} \bsnm{{Bagdonaite}}},
\bauthor{\binits{M.} \bsnm{{Dapr{\`a}}}},
\bauthor{\binits{P.} \bsnm{{Jansen}}},
\bauthor{\binits{H.L.} \bsnm{{Bethlem}}},
\bauthor{\binits{W.} \bsnm{{Ubachs}}},
\bauthor{\binits{S.} \bsnm{{Muller}}},
\bauthor{\binits{C.} \bsnm{{Henkel}}},
\bauthor{\binits{K.M.} \bsnm{{Menten}}},
\batitle{{Robust constraint on a drifting proton-to-electron mass ratio at
  $z=0.89$ from methanol observation at three radio telescopes}}.
\bjtitle{\prl}
\bvolume{111},
\bfpage{231101}
(\byear{2013}b)
\end{barticle}
\endbibitem

\bibitem[\protect\citeauthoryear{{Bagdonaite} et~al.}{2014a}]{Bagdonaite2014a}
\begin{barticle}
\bauthor{\binits{J.} \bsnm{{Bagdonaite}}},
\bauthor{\binits{W.} \bsnm{{Ubachs}}},
\bauthor{\binits{M.T.} \bsnm{{Murphy}}},
\bauthor{\binits{J.B.} \bsnm{{Whitmore}}},
\batitle{{Analysis of molecular hydrogen absorption toward QSO B0642$-$5038 for
  a varying proton-to-electron mass ratio}}.
\bjtitle{\apj}
\bvolume{782},
\bfpage{10}
(\byear{2014}a)
\end{barticle}
\endbibitem

\bibitem[\protect\citeauthoryear{{Bagdonaite} et~al.}{2014b}]{Bagdonaite2014b}
\begin{barticle}
\bauthor{\binits{J.} \bsnm{{Bagdonaite}}},
\bauthor{\binits{E.J.} \bsnm{{Salumbides}}},
\bauthor{\binits{S.P.} \bsnm{{Preval}}},
\bauthor{\binits{M.A.} \bsnm{{Barstow}}},
\bauthor{\binits{J.D.} \bsnm{{Barrow}}},
\bauthor{\binits{M.T.} \bsnm{{Murphy}}},
\bauthor{\binits{W.} \bsnm{{Ubachs}}},
\batitle{{Limits on a gravitational field dependence of the proton-electron
  mass ratio from H$_{2}$ in white dwarf stars}}.
\bjtitle{\prl}
\bvolume{113},
\bfpage{123002}
(\byear{2014}b)
\end{barticle}
\endbibitem

\bibitem[\protect\citeauthoryear{Bagdonaite et~al.}{2015}]{Bagdonaite2015}
\begin{barticle}
\bauthor{\binits{J.} \bsnm{Bagdonaite}},
\bauthor{\binits{W.} \bsnm{Ubachs}},
\bauthor{\binits{M.T.} \bsnm{Murphy}},
\bauthor{\binits{J.B.} \bsnm{Whitmore}},
\batitle{Constraint on a varying proton-electron mass ratio 1.5 billion years
  after the big bang}.
\bjtitle{\prl}
\bvolume{114},
\bfpage{071301}
(\byear{2015})
\end{barticle}
\endbibitem

\bibitem[\protect\citeauthoryear{Bahcall and Schmidt}{1967}]{Bahcall1967}
\begin{barticle}
\bauthor{\binits{J.N.} \bsnm{Bahcall}},
\bauthor{\binits{M.} \bsnm{Schmidt}},
\batitle{Does the fine-structure constant vary with cosmic time?}
\bjtitle{\prl}
\bvolume{19},
\bfpage{1294}
(\byear{1967})
\end{barticle}
\endbibitem

\bibitem[\protect\citeauthoryear{Bailly et~al.}{2010}]{Bailly2010}
\begin{barticle}
\bauthor{\binits{D.} \bsnm{Bailly}},
\bauthor{\binits{E.J.} \bsnm{Salumbides}},
\bauthor{\binits{M.} \bsnm{Vervloet}},
\bauthor{\binits{W.} \bsnm{Ubachs}},
\batitle{Accurate level energies in the {${EF}^{1}{\Sigma}^{+}_{g}$},
  {${GK}^{1}{\Sigma}^{+}_{g}$}, {${H}^{1}{\Sigma}^{+}_{g}$},
  ${B}^{1}{\Sigma}^{+}_{u}$, ${C}^{1}{\Pi}_{u}$, ${B'}^{1}{\Sigma}^{+}_{u}$,
  ${D}^{1}{\Pi}_{u}$, ${I}^{1}{\Pi}_{g}$, ${J}^{1}{\Delta}_{g}$ states of
  {H}$_{2}$}.
\bjtitle{Mol. Phys.}
\bvolume{108},
\bfpage{827}--\blpage{846}
(\byear{2010})
\end{barticle}
\endbibitem

\bibitem[\protect\citeauthoryear{{Balashev} et~al.}{2014}]{Balashev2014}
\begin{barticle}
\bauthor{\binits{S.A.} \bsnm{{Balashev}}},
\bauthor{\binits{V.V.} \bsnm{{Klimenko}}},
\bauthor{\binits{A.V.} \bsnm{{Ivanchik}}},
\bauthor{\binits{D.A.} \bsnm{{Varshalovich}}},
\bauthor{\binits{P.} \bsnm{{Petitjean}}},
\bauthor{\binits{P.} \bsnm{{Noterdaeme}}},
\batitle{{Molecular hydrogen absorption systems in Sloan Digital Sky Survey}}.
\bjtitle{\mnras}
\bvolume{440},
\bfpage{225}--\blpage{239}
(\byear{2014})
\end{barticle}
\endbibitem

\bibitem[\protect\citeauthoryear{Barrow et~al.}{2002}]{Barrow2002a}
\begin{barticle}
\bauthor{\binits{J.} \bsnm{Barrow}},
\bauthor{\binits{H.} \bsnm{Sandvik}},
\bauthor{\binits{J.} \bsnm{Magueijo}},
\batitle{Behavior of varying-$\alpha$ cosmologies}.
\bjtitle{\prd}
\bvolume{65},
\bfpage{063504}
(\byear{2002})
\end{barticle}
\endbibitem

\bibitem[\protect\citeauthoryear{Calmet and Fritzsch}{2002}]{Calmet2002}
\begin{barticle}
\bauthor{\binits{X.} \bsnm{Calmet}},
\bauthor{\binits{H.} \bsnm{Fritzsch}},
\batitle{The cosmological evolution of the nucleon mass and the electroweak
  coupling constants}.
\bjtitle{Eur. Phys. J. C}
\bvolume{24},
\bfpage{639}--\blpage{642}
(\byear{2002})
\end{barticle}
\endbibitem

\bibitem[\protect\citeauthoryear{{Carruthers}}{1970}]{Carruthers1970}
\begin{barticle}
\bauthor{\binits{G.R.} \bsnm{{Carruthers}}},
\batitle{Rocket observation of interstellar molecular hydrogen}.
\bjtitle{\apjl}
\bvolume{161},
\bfpage{81}
(\byear{1970})
\end{barticle}
\endbibitem

\bibitem[\protect\citeauthoryear{{Dapr\`{a}} et~al.}{2015}]{Dapra2015}
\begin{barticle}
\bauthor{\binits{M.} \bsnm{{Dapr\`{a}}}},
\bauthor{\binits{J.} \bsnm{{Bagdonaite}}},
\bauthor{\binits{M.T.} \bsnm{{Murphy}}},
\bauthor{\binits{W.} \bsnm{{Ubachs}}},
\batitle{{Constraint on a varying proton-to-electron mass ratio from molecular
  hydrogen absorption toward quasar SDSS J123714.60+064759.5}}.
\bjtitle{\mnras}
\bvolume{454},
\bfpage{489}--\blpage{506}
(\byear{2015})
\end{barticle}
\endbibitem

\bibitem[\protect\citeauthoryear{Dapr\`{a} et~al.}{2016}]{Dapra2016}
\begin{barticle}
\bauthor{\binits{M.} \bsnm{Dapr\`{a}}},
\bauthor{\binits{M.L.} \bsnm{Niu}},
\bauthor{\binits{E.J.} \bsnm{Salumbides}},
\bauthor{\binits{M.T.} \bsnm{Murphy}},
\bauthor{\binits{W.} \bsnm{Ubachs}},
\batitle{Constraint on a cosmological variation in the proton-to-electron mass
  ratio from electronic {CO} absorption}.
\bjtitle{\apj}
\bvolume{826},
\bfpage{192}
(\byear{2016})
\end{barticle}
\endbibitem

\bibitem[\protect\citeauthoryear{Dapr\'{a} et~al.}{2017a}]{Dapra2017a}
\begin{barticle}
\bauthor{\binits{M.} \bsnm{Dapr\'{a}}},
\bauthor{\binits{P.} \bsnm{Noterdaeme}},
\bauthor{\binits{M.} \bsnm{Vonk}},
\bauthor{\binits{M.T.} \bsnm{Murphy}},
\bauthor{\binits{W.} \bsnm{Ubachs}},
\batitle{Analysis of carbon monoxide absorption at $z_{abs}=2.5$ to constrain
  variation of the proton-to-electron mass ratio}.
\bjtitle{\mnras}
\bvolume{467},
\bfpage{3848}--\blpage{3855}
(\byear{2017}a)
\end{barticle}
\endbibitem

\bibitem[\protect\citeauthoryear{Dapr\'{a} et~al.}{2017b}]{Dapra2017}
\begin{barticle}
\bauthor{\binits{M.} \bsnm{Dapr\'{a}}},
\bauthor{\binits{M.} \bparticle{van~der} \bsnm{Laan}},
\bauthor{\binits{M.T.} \bsnm{Murphy}},
\bauthor{\binits{W.} \bsnm{Ubachs}},
\batitle{Constraint on a varying proton-to-electron mass ratio from {H}$_2$ and
  {HD} absorption at z$_{abs}$ = 2.34}.
\bjtitle{\mnras}
\bvolume{465},
\bfpage{4057}
(\byear{2017}b)
\end{barticle}
\endbibitem

\bibitem[\protect\citeauthoryear{Darling}{2004}]{Darling2004}
\begin{barticle}
\bauthor{\binits{J.} \bsnm{Darling}},
\batitle{A laboratory for constraining cosmic evolution of the fine-structure
  constant: Conjugate 18 centimeter {OH} lines toward {PKS} 1413+135 at $z =
  0.24671$}.
\bjtitle{Astrophys. J.}
\bvolume{612},
\bfpage{58}
(\byear{2004})
\end{barticle}
\endbibitem

\bibitem[\protect\citeauthoryear{Dumont and Webb}{2017}]{Dumont2017}
\begin{barticle}
\bauthor{\binits{V.} \bsnm{Dumont}},
\bauthor{\binits{J.K.} \bsnm{Webb}},
\batitle{Modelling long-range wavelength distortions in quasar absorption
  echelle spectra}.
\bjtitle{\mnras}
\bvolume{468},
\bfpage{1568}--\blpage{1574}
(\byear{2017})
\end{barticle}
\endbibitem

\bibitem[\protect\citeauthoryear{{Dzuba} et~al.}{1999}]{Dzuba1999}
\begin{barticle}
\bauthor{\binits{V.A.} \bsnm{{Dzuba}}},
\bauthor{\binits{V.V.} \bsnm{{Flambaum}}},
\bauthor{\binits{J.K.} \bsnm{{Webb}}},
\batitle{{Space-time variation of physical constants and relativistic
  corrections in atoms}}.
\bjtitle{\prl}
\bvolume{82},
\bfpage{888}--\blpage{891}
(\byear{1999})
\end{barticle}
\endbibitem

\bibitem[\protect\citeauthoryear{{Flambaum} and {Kozlov}}{2007}]{Flambaum2007}
\begin{barticle}
\bauthor{\binits{V.V.} \bsnm{{Flambaum}}},
\bauthor{\binits{M.G.} \bsnm{{Kozlov}}},
\batitle{{Limit on the cosmological variation of m$_{p}$/m$_{e}$ from the
  inversion spectrum of ammonia}}.
\bjtitle{\prl}
\bvolume{98},
\bfpage{240801}
(\byear{2007})
\end{barticle}
\endbibitem

\bibitem[\protect\citeauthoryear{{Flambaum} and {Shuryak}}{2008}]{Flambaum2008}
\begin{bchapter}
\bauthor{\binits{V.V.} \bsnm{{Flambaum}}},
\bauthor{\binits{E.V.} \bsnm{{Shuryak}}},
\bctitle{{How changing physical constants and violation of local position
  invariance may occur?}},
in \bbtitle{Nuclei and Mesoscopic Physics},
ed. by \beditor{\binits{P.} \bsnm{{Danielewicz}}},
\beditor{\binits{P.} \bsnm{{Piecuch}}},
\beditor{\binits{V.} \bsnm{{Zelevinsky}}}
\bsertitle{AIP Conference Series},
vol. \bseriesno{995},
\byear{2008},
pp. \bfpage{1}--\blpage{11}
\end{bchapter}
\endbibitem

\bibitem[\protect\citeauthoryear{France et~al.}{2016}]{France2016}
\begin{barticle}
\bauthor{\binits{K.C.} \bsnm{France}},
\bauthor{\binits{B.T.} \bsnm{Fleming}},
\bauthor{\binits{K.} \bsnm{Hoadley}},
\batitle{{CHISL}: the combined high-resolution and imaging spectrograph for the
  {LUVOIR} surveyor}.
\bjtitle{J. Astron. Telesc. Instr. Systems}
\bvolume{2},
\bfpage{041203}
(\byear{2016})
\end{barticle}
\endbibitem

\bibitem[\protect\citeauthoryear{{Griest} et~al.}{2010}]{Griest2010}
\begin{barticle}
\bauthor{\binits{K.} \bsnm{{Griest}}},
\bauthor{\binits{J.B.} \bsnm{{Whitmore}}},
\bauthor{\binits{A.M.} \bsnm{{Wolfe}}},
\bauthor{\binits{J.X.} \bsnm{{Prochaska}}},
\bauthor{\binits{J.C.} \bsnm{{Howk}}},
\bauthor{\binits{G.W.} \bsnm{{Marcy}}},
\batitle{{Wavelength accuracy of the Keck HIRES spectrograph and measuring
  changes in the fine structure constant}}.
\bjtitle{\apj}
\bvolume{708},
\bfpage{158}--\blpage{170}
(\byear{2010})
\end{barticle}
\endbibitem

\bibitem[\protect\citeauthoryear{{Henkel} et~al.}{2009}]{Henkel2009}
\begin{barticle}
\bauthor{\binits{C.} \bsnm{{Henkel}}},
\bauthor{\binits{K.M.} \bsnm{{Menten}}},
\bauthor{\binits{M.T.} \bsnm{{Murphy}}},
\bauthor{\binits{N.} \bsnm{{Jethava}}},
\bauthor{\binits{V.V.} \bsnm{{Flambaum}}},
\bauthor{\binits{J.A.} \bsnm{{Braatz}}},
\bauthor{\binits{S.} \bsnm{{Muller}}},
\bauthor{\binits{J.} \bsnm{{Ott}}},
\bauthor{\binits{R.Q.} \bsnm{{Mao}}},
\batitle{{The density, the cosmic microwave background, and the
  proton-to-electron mass ratio in a cloud at redshift 0.9}}.
\bjtitle{\aap}
\bvolume{500},
\bfpage{725}--\blpage{734}
(\byear{2009})
\end{barticle}
\endbibitem

\bibitem[\protect\citeauthoryear{{Ivanchik} et~al.}{2010}]{Ivanchik2010}
\begin{barticle}
\bauthor{\binits{A.V.} \bsnm{{Ivanchik}}},
\bauthor{\binits{P.} \bsnm{{Petitjean}}},
\bauthor{\binits{S.A.} \bsnm{{Balashev}}},
\bauthor{\binits{R.} \bsnm{{Srianand}}},
\bauthor{\binits{D.A.} \bsnm{{Varshalovich}}},
\bauthor{\binits{C.} \bsnm{{Ledoux}}},
\bauthor{\binits{P.} \bsnm{{Noterdaeme}}},
\batitle{{HD molecules at high redshift: the absorption system at $z = 2.3377$
  towards Q1232$+$082}}.
\bjtitle{\mnras}
\bvolume{404},
\bfpage{1583}--\blpage{1590}
(\byear{2010})
\end{barticle}
\endbibitem

\bibitem[\protect\citeauthoryear{Ivanchik et~al.}{2005}]{Ivanchik2005}
\begin{barticle}
\bauthor{\binits{A.} \bsnm{Ivanchik}},
\bauthor{\binits{P.} \bsnm{Petitjean}},
\bauthor{\binits{D.} \bsnm{Varshalovich}},
\bauthor{\binits{B.} \bsnm{Aracil}},
\bauthor{\binits{R.} \bsnm{Srianand}},
\bauthor{\binits{H.} \bsnm{Chand}},
\bauthor{\binits{C.} \bsnm{Ledoux}},
\bauthor{\binits{P.} \bsnm{Boiss{\'e}}},
\batitle{{A new constraint on the time dependence of the proton-to-electron
  mass ratio. Analysis of the Q0347$-$383 and Q0405$-$443 spectra}}.
\bjtitle{\aap}
\bvolume{440},
\bfpage{45}--\blpage{52}
(\byear{2005})
\end{barticle}
\endbibitem

\bibitem[\protect\citeauthoryear{{Ivanov} et~al.}{2008}]{Ivanov2008a}
\begin{barticle}
\bauthor{\binits{T.I.} \bsnm{{Ivanov}}},
\bauthor{\binits{M.} \bsnm{{Roudjane}}},
\bauthor{\binits{M.O.} \bsnm{{Vieitez}}},
\bauthor{\binits{C.A.} \bsnm{{de Lange}}},
\bauthor{\binits{W.-{\"U}.L.} \bsnm{{Tchang-Brillet}}},
\bauthor{\binits{W.} \bsnm{{Ubachs}}},
\batitle{{HD as a probe for detecting mass variation on a cosmological time
  scale}}.
\bjtitle{\prl}
\bvolume{100},
\bfpage{093007}
(\byear{2008})
\end{barticle}
\endbibitem

\bibitem[\protect\citeauthoryear{{Jansen} et~al.}{2014}]{Jansen2014}
\begin{barticle}
\bauthor{\binits{P.} \bsnm{{Jansen}}},
\bauthor{\binits{H.L.} \bsnm{{Bethlem}}},
\bauthor{\binits{W.} \bsnm{{Ubachs}}},
\batitle{{Perspective: Tipping the scales: Search for drifting constants from
  molecular spectra}}.
\bjtitle{\jcp}
\bvolume{140},
\bfpage{010901}
(\byear{2014})
\end{barticle}
\endbibitem

\bibitem[\protect\citeauthoryear{{Jansen} et~al.}{2011}]{Jansen2011}
\begin{barticle}
\bauthor{\binits{P.} \bsnm{{Jansen}}},
\bauthor{\binits{L.-H.} \bsnm{{Xu}}},
\bauthor{\binits{I.} \bsnm{{Kleiner}}},
\bauthor{\binits{W.} \bsnm{{Ubachs}}},
\bauthor{\binits{H.L.} \bsnm{{Bethlem}}},
\batitle{{Methanol as a sensitive probe for spatial and temporal variations of
  the proton-to-electron mass ratio}}.
\bjtitle{\prl}
\bvolume{106},
\bfpage{100801}
(\byear{2011})
\end{barticle}
\endbibitem

\bibitem[\protect\citeauthoryear{{Kanekar}}{2011}]{Kanekar2011}
\begin{barticle}
\bauthor{\binits{N.} \bsnm{{Kanekar}}},
\batitle{{Constraining changes in the proton-electron mass ratio with inversion
  and rotational Lines}}.
\bjtitle{\apjl}
\bvolume{728},
\bfpage{12}
(\byear{2011})
\end{barticle}
\endbibitem

\bibitem[\protect\citeauthoryear{Kanekar and Chengalur}{2002}]{Kanekar2002}
\begin{barticle}
\bauthor{\binits{N.} \bsnm{Kanekar}},
\bauthor{\binits{J.N.} \bsnm{Chengalur}},
\batitle{Molecular gas at intermediate redshifts}.
\bjtitle{Astron. Astroph.}
\bvolume{381},
\bfpage{73}--\blpage{76}
(\byear{2002})
\end{barticle}
\endbibitem

\bibitem[\protect\citeauthoryear{{Kanekar} et~al.}{2005}]{Kanekar2005}
\begin{barticle}
\bauthor{\binits{N.} \bsnm{{Kanekar}}},
\bauthor{\binits{C.L.} \bsnm{{Carilli}}},
\bauthor{\binits{G.I.} \bsnm{{Langston}}},
\bauthor{\binits{G.} \bsnm{{Rocha}}},
\bauthor{\binits{F.} \bsnm{{Combes}}},
\bauthor{\binits{R.} \bsnm{{Subrahmanyan}}},
\bauthor{\binits{J.T.} \bsnm{{Stocke}}},
\bauthor{\binits{K.M.} \bsnm{{Menten}}},
\bauthor{\binits{F.H.} \bsnm{{Briggs}}},
\bauthor{\binits{T.} \bsnm{{Wiklind}}},
\batitle{{Constraints on changes in fundamental constants from a cosmologically
  distant OH absorber or emitter}}.
\bjtitle{\prl}
\bvolume{95},
\bfpage{261301}
(\byear{2005})
\end{barticle}
\endbibitem

\bibitem[\protect\citeauthoryear{Kanekar et~al.}{2015}]{Kanekar2015}
\begin{barticle}
\bauthor{\binits{N.} \bsnm{Kanekar}},
\bauthor{\binits{W.} \bsnm{Ubachs}},
\bauthor{\binits{K.M.} \bsnm{Menten}},
\bauthor{\binits{J.} \bsnm{Bagdonaite}},
\bauthor{\binits{A.} \bsnm{Brunthaler}},
\bauthor{\binits{C.} \bsnm{Henkel}},
\bauthor{\binits{S.} \bsnm{Muller}},
\bauthor{\binits{H.L.} \bsnm{Bethlem}},
\bauthor{\binits{M.} \bsnm{Dapr\`{a}}},
\batitle{{Constraints on changes in the proton-electron mass ratio using
  methanol lines}}.
\bjtitle{\mnras}
\bvolume{448},
\bfpage{104}--\blpage{108}
(\byear{2015})
\end{barticle}
\endbibitem

\bibitem[\protect\citeauthoryear{{Khoury} and {Weltman}}{2004}]{Khoury2004}
\begin{barticle}
\bauthor{\binits{J.} \bsnm{{Khoury}}},
\bauthor{\binits{A.} \bsnm{{Weltman}}},
\batitle{{Chameleon fields: Awaiting surprises for tests of gravity in space}}.
\bjtitle{\prl}
\bvolume{93},
\bfpage{171104}
(\byear{2004})
\end{barticle}
\endbibitem

\bibitem[\protect\citeauthoryear{{King} et~al.}{2008}]{King2008}
\begin{barticle}
\bauthor{\binits{J.A.} \bsnm{{King}}},
\bauthor{\binits{J.K.} \bsnm{{Webb}}},
\bauthor{\binits{M.T.} \bsnm{{Murphy}}},
\bauthor{\binits{R.F.} \bsnm{{Carswell}}},
\batitle{{Stringent null constraint on cosmological evolution of the
  proton-to-electron mass ratio}}.
\bjtitle{\prl}
\bvolume{101},
\bfpage{251304}
(\byear{2008})
\end{barticle}
\endbibitem

\bibitem[\protect\citeauthoryear{{King} et~al.}{2011}]{King2011}
\begin{barticle}
\bauthor{\binits{J.A.} \bsnm{{King}}},
\bauthor{\binits{M.T.} \bsnm{{Murphy}}},
\bauthor{\binits{W.} \bsnm{{Ubachs}}},
\bauthor{\binits{J.K.} \bsnm{{Webb}}},
\batitle{{New constraint on cosmological variation of the proton-to-electron
  mass ratio from Q0528$-$250}}.
\bjtitle{\mnras}
\bvolume{417},
\bfpage{3010}--\blpage{3024}
(\byear{2011})
\end{barticle}
\endbibitem

\bibitem[\protect\citeauthoryear{Kozlov and Levshakov}{2013}]{Kozlov2013b}
\begin{barticle}
\bauthor{\binits{M.G.} \bsnm{Kozlov}},
\bauthor{\binits{S.A.} \bsnm{Levshakov}},
\batitle{Microwave and submillimeter molecular transitions and their dependence
  on fundamental constants}.
\bjtitle{Ann. der Phys.}
\bvolume{525},
\bfpage{452}--\blpage{471}
(\byear{2013})
\end{barticle}
\endbibitem

\bibitem[\protect\citeauthoryear{Langacker et~al.}{2002}]{Langacker2002}
\begin{barticle}
\bauthor{\binits{P.} \bsnm{Langacker}},
\bauthor{\binits{G.} \bsnm{Segr{\'e}}},
\bauthor{\binits{M.J.} \bsnm{Strassler}},
\batitle{Implications of gauge unification for time variation of the fine
  structure constant}.
\bjtitle{Phys. Lett. B}
\bvolume{528},
\bfpage{121}--\blpage{128}
(\byear{2002})
\end{barticle}
\endbibitem

\bibitem[\protect\citeauthoryear{{Levshakov} et~al.}{2011}]{Levshakov2011}
\begin{barticle}
\bauthor{\binits{S.A.} \bsnm{{Levshakov}}},
\bauthor{\binits{M.G.} \bsnm{{Kozlov}}},
\bauthor{\binits{D.} \bsnm{{Reimers}}},
\batitle{{Methanol as a tracer of fundamental constants}}.
\bjtitle{\apj}
\bvolume{738},
\bfpage{26}
(\byear{2011})
\end{barticle}
\endbibitem

\bibitem[\protect\citeauthoryear{{Levshakov} et~al.}{2012}]{Levshakov2012}
\begin{barticle}
\bauthor{\binits{S.A.} \bsnm{{Levshakov}}},
\bauthor{\binits{F.} \bsnm{{Combes}}},
\bauthor{\binits{F.} \bsnm{{Boone}}},
\bauthor{\binits{I.I.} \bsnm{{Agafonova}}},
\bauthor{\binits{D.} \bsnm{{Reimers}}},
\bauthor{\binits{M.G.} \bsnm{{Kozlov}}},
\batitle{{An upper limit to the variation in the fundamental constants at
  redshift $z = 5.2$}}.
\bjtitle{\aap}
\bvolume{540},
\bfpage{9}
(\byear{2012})
\end{barticle}
\endbibitem

\bibitem[\protect\citeauthoryear{{Malec} et~al.}{2010}]{Malec2010}
\begin{barticle}
\bauthor{\binits{A.L.} \bsnm{{Malec}}},
\bauthor{\binits{R.} \bsnm{{Buning}}},
\bauthor{\binits{M.T.} \bsnm{{Murphy}}},
\bauthor{\binits{N.} \bsnm{{Milutinovic}}},
\bauthor{\binits{S.L.} \bsnm{{Ellison}}},
\bauthor{\binits{J.X.} \bsnm{{Prochaska}}},
\bauthor{\binits{L.} \bsnm{{Kaper}}},
\bauthor{\binits{J.} \bsnm{{Tumlinson}}},
\bauthor{\binits{R.F.} \bsnm{{Carswell}}},
\bauthor{\binits{W.} \bsnm{{Ubachs}}},
\batitle{{Keck telescope constraint on cosmological variation of the
  proton-to-electron mass ratio}}.
\bjtitle{\mnras}
\bvolume{403},
\bfpage{1541}--\blpage{1555}
(\byear{2010})
\end{barticle}
\endbibitem

\bibitem[\protect\citeauthoryear{{Meshkov} et~al.}{2006}]{Meshkov2006}
\begin{barticle}
\bauthor{\binits{V.V.} \bsnm{{Meshkov}}},
\bauthor{\binits{A.V.} \bsnm{{Stolyarov}}},
\bauthor{\binits{A.V.} \bsnm{{Ivanchik}}},
\bauthor{\binits{D.A.} \bsnm{{Varshalovich}}},
\batitle{{Ab initio nonadiabatic calculation of the sensitivity coefficients
  for the $X^1\Sigma_g^+$ {$\rightarrow$} $B^1\Sigma_u^+$ ; $C^1\Pi_u$ lines of
  H$_2$ to the proton-to-electron mass ratio}}.
\bjtitle{Sov. J. Exp. Theor. Phys.}
\bvolume{83},
\bfpage{303}--\blpage{307}
(\byear{2006})
\end{barticle}
\endbibitem

\bibitem[\protect\citeauthoryear{{Molaro} et~al.}{2008}]{Molaro2008}
\begin{barticle}
\bauthor{\binits{P.} \bsnm{{Molaro}}},
\bauthor{\binits{S.A.} \bsnm{{Levshakov}}},
\bauthor{\binits{S.} \bsnm{{Monai}}},
\bauthor{\binits{M.} \bsnm{{Centuri{\'o}n}}},
\bauthor{\binits{P.} \bsnm{{Bonifacio}}},
\bauthor{\binits{S.} \bsnm{{D'Odorico}}},
\bauthor{\binits{L.} \bsnm{{Monaco}}},
\batitle{{UVES radial velocity accuracy from asteroid observations. I.
  Implications for fine structure constant variability}}.
\bjtitle{\aap}
\bvolume{481},
\bfpage{559}--\blpage{569}
(\byear{2008})
\end{barticle}
\endbibitem

\bibitem[\protect\citeauthoryear{{Muller} et~al.}{2011}]{Muller2011}
\begin{barticle}
\bauthor{\binits{S.} \bsnm{{Muller}}},
\bauthor{\binits{A.} \bsnm{{Beelen}}},
\bauthor{\binits{M.} \bsnm{{Gu{\'e}lin}}},
\bauthor{\binits{S.} \bsnm{{Aalto}}},
\bauthor{\binits{J.H.} \bsnm{{Black}}},
\bauthor{\binits{F.} \bsnm{{Combes}}},
\bauthor{\binits{S.J.} \bsnm{{Curran}}},
\bauthor{\binits{P.} \bsnm{{Theule}}},
\bauthor{\binits{S.N.} \bsnm{{Longmore}}},
\batitle{{Molecules at $z = 0.89$. A 4-mm-rest-frame absorption-line survey
  toward PKS 1830$-$211}}.
\bjtitle{\aap}
\bvolume{535},
\bfpage{103}
(\byear{2011})
\end{barticle}
\endbibitem

\bibitem[\protect\citeauthoryear{{Murphy} et~al.}{2008}]{Murphy2008a}
\begin{barticle}
\bauthor{\binits{M.T.} \bsnm{{Murphy}}},
\bauthor{\binits{V.V.} \bsnm{{Flambaum}}},
\bauthor{\binits{S.} \bsnm{{Muller}}},
\bauthor{\binits{C.} \bsnm{{Henkel}}},
\batitle{{Strong limit on a variable proton-to-electron mass ratio from
  molecules in the distant universe}}.
\bjtitle{Science}
\bvolume{320},
\bfpage{1611}--\blpage{1613}
(\byear{2008})
\end{barticle}
\endbibitem

\bibitem[\protect\citeauthoryear{Noterdaeme et~al.}{2008}]{Noterdaeme2008}
\begin{barticle}
\bauthor{\binits{P.} \bsnm{Noterdaeme}},
\bauthor{\binits{C.} \bsnm{Ledoux}},
\bauthor{\binits{P.} \bsnm{Petitjean}},
\bauthor{\binits{R.} \bsnm{Srianand}},
\batitle{{Molecular hydrogen in high-redshift damped Lyman-{$\alpha$} systems:
  the VLT/UVES database}}.
\bjtitle{\aap}
\bvolume{481},
\bfpage{327}--\blpage{336}
(\byear{2008})
\end{barticle}
\endbibitem

\bibitem[\protect\citeauthoryear{Noterdaeme et~al.}{2010}]{Noterdaeme2010}
\begin{barticle}
\bauthor{\binits{P.} \bsnm{Noterdaeme}},
\bauthor{\binits{P.} \bsnm{Petitjean}},
\bauthor{\binits{C.} \bsnm{Ledoux}},
\bauthor{\binits{S.} \bsnm{L{\'o}pez}},
\bauthor{\binits{R.} \bsnm{Srianand}},
\bauthor{\binits{S.D.} \bsnm{Vergani}},
\batitle{{A translucent interstellar cloud at $z = 2.69$. CO, H$_{2}$, and HD
  in the line-of-sight to SDSS J123714.60+064759.5}}.
\bjtitle{\aap}
\bvolume{523},
\bfpage{80}
(\byear{2010})
\end{barticle}
\endbibitem

\bibitem[\protect\citeauthoryear{Noterdaeme et~al.}{2011}]{Noterdaeme2011}
\begin{barticle}
\bauthor{\binits{P.} \bsnm{Noterdaeme}},
\bauthor{\binits{P.} \bsnm{Petitjean}},
\bauthor{\binits{R.} \bsnm{Srianand}},
\bauthor{\binits{C.} \bsnm{Ledoux}},
\bauthor{\binits{S.} \bsnm{L{\'o}pez}},
\batitle{{The evolution of the cosmic microwave background temperature.
  Measurements of T$_{\textrm{CMB}}$ at high redshift from carbon monoxide
  excitation}}.
\bjtitle{\aap}
\bvolume{526},
\bfpage{7}
(\byear{2011})
\end{barticle}
\endbibitem

\bibitem[\protect\citeauthoryear{Noterdaeme et~al.}{2017}]{Noterdaeme2017}
\begin{barticle}
\bauthor{\binits{P.} \bsnm{Noterdaeme}},
\bauthor{\binits{J.-K.} \bsnm{Krogager}},
\bauthor{\binits{S.} \bsnm{Balashev}},
\bauthor{\binits{J.} \bsnm{Ge}},
\bauthor{\binits{N.} \bsnm{Gupta}},
\bauthor{\binits{T.} \bsnm{Kr\"{u}hler}},
\bauthor{\binits{C.} \bsnm{Ledoux}},
\bauthor{\binits{M.T.} \bsnm{Murphy}},
\bauthor{\binits{I.} \bsnm{P\^{a}ris}},
\bauthor{\binits{P.} \bsnm{Petitjean}},
\bauthor{\binits{H.} \bsnm{Rahmani}},
\bauthor{\binits{R.} \bsnm{Srianand}},
\bauthor{\binits{W.} \bsnm{Ubachs}},
\batitle{Discovery of a {P}erseus-like cloud in the early {U}niverse -
  {HI}-to-{H$_2$} transition, carbon monoxide and small dust grains at
  $z_{abs}\approx 2.53$ towards the quasar {J}0000+0048}.
\bjtitle{\aap}
\bvolume{597},
\bfpage{82}
(\byear{2017})
\end{barticle}
\endbibitem

\bibitem[\protect\citeauthoryear{{Petitjean} et~al.}{2002}]{Petitjean2002}
\begin{barticle}
\bauthor{\binits{P.} \bsnm{{Petitjean}}},
\bauthor{\binits{R.} \bsnm{{Srianand}}},
\bauthor{\binits{C.} \bsnm{{Ledoux}}},
\batitle{{Molecular hydrogen at $z_{abs} = 1.973$ toward Q0013$-$004: dust
  depletion pattern in damped Lyman {$\alpha$} systems}}.
\bjtitle{\mnras}
\bvolume{332},
\bfpage{383}--\blpage{391}
(\byear{2002})
\end{barticle}
\endbibitem

\bibitem[\protect\citeauthoryear{Philip et~al.}{2004}]{Philip2004}
\begin{barticle}
\bauthor{\binits{J.} \bsnm{Philip}},
\bauthor{\binits{J.P.} \bsnm{Sprengers}},
\bauthor{\binits{T.} \bsnm{Pielage}},
\bauthor{\binits{C.A.} \bparticle{de} \bsnm{Lange}},
\bauthor{\binits{W.} \bsnm{Ubachs}},
\bauthor{\binits{E.} \bsnm{Reinhold}},
\batitle{Highly accurate transition frequencies in the {H}$_2$ {L}yman and
  {W}erner absorption bands}.
\bjtitle{Can. J. Chem.}
\bvolume{82},
\bfpage{713}--\blpage{722}
(\byear{2004})
\end{barticle}
\endbibitem

\bibitem[\protect\citeauthoryear{Rahmani et~al.}{2013}]{Rahmani2013}
\begin{barticle}
\bauthor{\binits{H.} \bsnm{Rahmani}},
\bauthor{\binits{M.} \bsnm{Wendt}},
\bauthor{\binits{R.} \bsnm{Srianand}},
\bauthor{\binits{P.} \bsnm{Noterdaeme}},
\bauthor{\binits{P.} \bsnm{Petitjean}},
\bauthor{\binits{P.} \bsnm{Molaro}},
\bauthor{\binits{J.B.} \bsnm{Whitmore}},
\bauthor{\binits{M.T.} \bsnm{Murphy}},
\bauthor{\binits{M.} \bsnm{Centurion}},
\bauthor{\binits{H.} \bsnm{Fathivavsari}},
\bauthor{\binits{S.} \bsnm{D'Odorico}},
\bauthor{\binits{T.M.} \bsnm{Evans}},
\bauthor{\binits{S.A.} \bsnm{Levshakov}},
\bauthor{\binits{S.} \bsnm{Lopez}},
\bauthor{\binits{C.J.A.P.} \bsnm{Martins}},
\bauthor{\binits{D.} \bsnm{Reimers}},
\bauthor{\binits{G.} \bsnm{Vladilo}},
\batitle{{The UVES large program for testing fundamental physics - II.
  Constraints on a change in {$\mu$} towards quasar HE 0027$-$1836}}.
\bjtitle{\mnras}
\bvolume{435},
\bfpage{861}--\blpage{878}
(\byear{2013})
\end{barticle}
\endbibitem

\bibitem[\protect\citeauthoryear{Rauch}{1998}]{Rauch1998}
\begin{barticle}
\bauthor{\binits{M.} \bsnm{Rauch}},
\batitle{The {L}yman alpha forest in the spectra of quasistellar objects}.
\bjtitle{Ann. Rev. Astron. and Astroph.}
\bvolume{36},
\bfpage{267}--\blpage{316}
(\byear{1998})
\end{barticle}
\endbibitem

\bibitem[\protect\citeauthoryear{{Reinhold} et~al.}{2006}]{Reinhold2006}
\begin{barticle}
\bauthor{\binits{E.} \bsnm{{Reinhold}}},
\bauthor{\binits{R.} \bsnm{{Buning}}},
\bauthor{\binits{U.} \bsnm{{Hollenstein}}},
\bauthor{\binits{A.} \bsnm{{Ivanchik}}},
\bauthor{\binits{P.} \bsnm{{Petitjean}}},
\bauthor{\binits{W.} \bsnm{{Ubachs}}},
\batitle{{Indication of a cosmological variation of the proton-electron mass
  ratio based on laboratory measurement and reanalysis of H$_{2}$ spectra}}.
\bjtitle{\prl}
\bvolume{96},
\bfpage{151101}
(\byear{2006})
\end{barticle}
\endbibitem

\bibitem[\protect\citeauthoryear{{Salumbides} et~al.}{2012}]{Salumbides2012}
\begin{barticle}
\bauthor{\binits{E.J.} \bsnm{{Salumbides}}},
\bauthor{\binits{M.L.} \bsnm{{Niu}}},
\bauthor{\binits{J.} \bsnm{{Bagdonaite}}},
\bauthor{\binits{N.} \bsnm{{de Oliveira}}},
\bauthor{\binits{D.} \bsnm{{Joyeux}}},
\bauthor{\binits{L.} \bsnm{{Nahon}}},
\bauthor{\binits{W.} \bsnm{{Ubachs}}},
\batitle{{CO A-X system for constraining cosmological drift of the
  proton--electron mass ratio}}.
\bjtitle{\pra}
\bvolume{86},
\bfpage{022510}
(\byear{2012})
\end{barticle}
\endbibitem

\bibitem[\protect\citeauthoryear{Salumbides et~al.}{2015}]{Salumbides2015}
\begin{barticle}
\bauthor{\binits{E.J.} \bsnm{Salumbides}},
\bauthor{\binits{J.} \bsnm{Bagdonaite}},
\bauthor{\binits{H.} \bsnm{Abgrall}},
\bauthor{\binits{E.} \bsnm{Roueff}},
\bauthor{\binits{W.} \bsnm{Ubachs}},
\batitle{{H$_2$} {L}yman and {W}erner band lines and their sensitivity for a
  variation of the proton–-electron mass ratio in the gravitational potential
  of white dwarfs}.
\bjtitle{\mnras}
\bvolume{450},
\bfpage{1237}--\blpage{1245}
(\byear{2015})
\end{barticle}
\endbibitem

\bibitem[\protect\citeauthoryear{{Savedoff}}{1956}]{Savedoff1956}
\begin{barticle}
\bauthor{\binits{M.P.} \bsnm{{Savedoff}}},
\batitle{{Physical constants in extra-galactic nebul{\ae}}}.
\bjtitle{Nature}
\bvolume{178},
\bfpage{688}--\blpage{689}
(\byear{1956})
\end{barticle}
\endbibitem

\bibitem[\protect\citeauthoryear{Srianand et~al.}{2005}]{Srianand2005}
\begin{barticle}
\bauthor{\binits{R.} \bsnm{Srianand}},
\bauthor{\binits{P.} \bsnm{Petitjean}},
\bauthor{\binits{C.} \bsnm{Ledoux}},
\bauthor{\binits{G.} \bsnm{Ferland}},
\bauthor{\binits{G.} \bsnm{Shaw}},
\batitle{{The VLT-UVES survey for molecular hydrogen in high-redshift damped
  Lyman {$\alpha$} systems: physical conditions in the neutral gas}}.
\bjtitle{\mnras}
\bvolume{362},
\bfpage{549}--\blpage{568}
(\byear{2005})
\end{barticle}
\endbibitem

\bibitem[\protect\citeauthoryear{{Thompson} et~al.}{2009}]{Thompson2009}
\begin{barticle}
\bauthor{\binits{R.I.} \bsnm{{Thompson}}},
\bauthor{\binits{J.} \bsnm{{Bechtold}}},
\bauthor{\binits{J.H.} \bsnm{{Black}}},
\bauthor{\binits{D.} \bsnm{{Eisenstein}}},
\bauthor{\binits{X.} \bsnm{{Fan}}},
\bauthor{\binits{R.C.} \bsnm{{Kennicutt}}},
\bauthor{\binits{C.} \bsnm{{Martins}}},
\bauthor{\binits{J.X.} \bsnm{{Prochaska}}},
\bauthor{\binits{Y.L.} \bsnm{{Shirley}}},
\batitle{{An observational determination of the proton-to-electron mass ratio
  in the early universe}}.
\bjtitle{\apj}
\bvolume{703},
\bfpage{1648}--\blpage{1662}
(\byear{2009})
\end{barticle}
\endbibitem

\bibitem[\protect\citeauthoryear{Ubachs and Reinhold}{2004}]{Ubachs2004}
\begin{barticle}
\bauthor{\binits{W.} \bsnm{Ubachs}},
\bauthor{\binits{E.} \bsnm{Reinhold}},
\batitle{Highly accurate {H$_2$} {L}yman and {W}erner band laboratory
  measurements and an improved constraint on a cosmological variation of the
  proton-to-electron mass ratio}.
\bjtitle{\prl}
\bvolume{92},
\bfpage{101302}
(\byear{2004})
\end{barticle}
\endbibitem

\bibitem[\protect\citeauthoryear{{Ubachs} et~al.}{2007}]{Ubachs2007}
\begin{barticle}
\bauthor{\binits{W.} \bsnm{{Ubachs}}},
\bauthor{\binits{R.} \bsnm{{Buning}}},
\bauthor{\binits{K.S.E.} \bsnm{{Eikema}}},
\bauthor{\binits{E.} \bsnm{{Reinhold}}},
\batitle{{On a possible variation of the proton-to-electron mass ratio: H$_{2}$
  spectra in the line of sight of high-redshift quasars and in the
  laboratory}}.
\bjtitle{\jms}
\bvolume{241},
\bfpage{155}--\blpage{179}
(\byear{2007})
\end{barticle}
\endbibitem

\bibitem[\protect\citeauthoryear{Ubachs et~al.}{2016}]{Ubachs2016}
\begin{barticle}
\bauthor{\binits{W.} \bsnm{Ubachs}},
\bauthor{\binits{J.} \bsnm{Bagdonaite}},
\bauthor{\binits{E.J.} \bsnm{Salumbides}},
\bauthor{\binits{M.T.} \bsnm{Murphy}},
\bauthor{\binits{L.} \bsnm{Kaper}},
\batitle{\textit{Colloquium} : Search for a drifting proton-electron mass ratio
  from {H}$_2$}.
\bjtitle{Rev. Mod. Phys.}
\bvolume{88},
\bfpage{021003}
(\byear{2016})
\end{barticle}
\endbibitem

\bibitem[\protect\citeauthoryear{{Uzan}}{2011}]{Uzan2011}
\begin{barticle}
\bauthor{\binits{J.-P.} \bsnm{{Uzan}}},
\batitle{{Varying constants, gravitation and cosmology}}.
\bjtitle{\lrr}
\bvolume{14},
\bfpage{2}
(\byear{2011})
\end{barticle}
\endbibitem

\bibitem[\protect\citeauthoryear{{van Weerdenburg}
  et~al.}{2011}]{Weerdenburg2011}
\begin{barticle}
\bauthor{\binits{F.} \bsnm{{van Weerdenburg}}},
\bauthor{\binits{M.T.} \bsnm{{Murphy}}},
\bauthor{\binits{A.L.} \bsnm{{Malec}}},
\bauthor{\binits{L.} \bsnm{{Kaper}}},
\bauthor{\binits{W.} \bsnm{{Ubachs}}},
\batitle{{First constraint on cosmological variation of the proton-to-electron
  mass ratio from two independent telescopes}}.
\bjtitle{\prl}
\bvolume{106},
\bfpage{180802}
(\byear{2011})
\end{barticle}
\endbibitem

\bibitem[\protect\citeauthoryear{{Varshalovich} and
  {Levshakov}}{1993}]{Varshalovich1993}
\begin{barticle}
\bauthor{\binits{D.A.} \bsnm{{Varshalovich}}},
\bauthor{\binits{S.A.} \bsnm{{Levshakov}}},
\batitle{{On a time dependence of physical constants.}}
\bjtitle{Sov. J. Exp. Theor. Phys.}
\bvolume{58},
\bfpage{237}--\blpage{240}
(\byear{1993})
\end{barticle}
\endbibitem

\bibitem[\protect\citeauthoryear{{Varshalovich}
  et~al.}{2001}]{Varshalovich2001}
\begin{barticle}
\bauthor{\binits{D.A.} \bsnm{{Varshalovich}}},
\bauthor{\binits{A.V.} \bsnm{{Ivanchik}}},
\bauthor{\binits{P.} \bsnm{{Petitjean}}},
\bauthor{\binits{R.} \bsnm{{Srianand}}},
\bauthor{\binits{C.} \bsnm{{Ledoux}}},
\batitle{{HD molecular lines in an absorption system at redshift $z =
  2.3377$}}.
\bjtitle{Astron. Lett.}
\bvolume{27},
\bfpage{683}--\blpage{685}
(\byear{2001})
\end{barticle}
\endbibitem

\bibitem[\protect\citeauthoryear{Webb et~al.}{2001}]{Webb2001}
\begin{barticle}
\bauthor{\binits{J.K.} \bsnm{Webb}},
\bauthor{\binits{M.T.} \bsnm{Murphy}},
\bauthor{\binits{V.V.} \bsnm{Flambaum}},
\bauthor{\binits{V.A.} \bsnm{Dzuba}},
\bauthor{\binits{J.D.} \bsnm{Barrow}},
\bauthor{\binits{C.W.} \bsnm{Churchill}},
\bauthor{\binits{J.X.} \bsnm{Prochaska}},
\bauthor{\binits{A.M.} \bsnm{Wolfe}},
\batitle{Further evidence for cosmological evolution of the fine structure
  constant}.
\bjtitle{Phys. Rev. Lett.}
\bvolume{87},
\bfpage{091301}
(\byear{2001})
\end{barticle}
\endbibitem

\bibitem[\protect\citeauthoryear{{Webb} et~al.}{2011}]{Webb2011}
\begin{barticle}
\bauthor{\binits{J.K.} \bsnm{{Webb}}},
\bauthor{\binits{J.A.} \bsnm{{King}}},
\bauthor{\binits{M.T.} \bsnm{{Murphy}}},
\bauthor{\binits{V.V.} \bsnm{{Flambaum}}},
\bauthor{\binits{R.F.} \bsnm{{Carswell}}},
\bauthor{\binits{M.B.} \bsnm{{Bainbridge}}},
\batitle{{Indications of a spatial variation of the fine structure constant}}.
\bjtitle{\prl}
\bvolume{107},
\bfpage{191101}
(\byear{2011})
\end{barticle}
\endbibitem

\bibitem[\protect\citeauthoryear{Wendt and Molaro}{2012}]{Wendt2012}
\begin{barticle}
\bauthor{\binits{M.} \bsnm{Wendt}},
\bauthor{\binits{P.} \bsnm{Molaro}},
\batitle{{QSO 0347$-$383 and the invariance of m$_{p}$/m$_{e}$ in the course of
  cosmic time}}.
\bjtitle{\aap}
\bvolume{541},
\bfpage{69}
(\byear{2012})
\end{barticle}
\endbibitem

\bibitem[\protect\citeauthoryear{{Whitmore} and {Murphy}}{2015}]{Whitmore2015}
\begin{barticle}
\bauthor{\binits{J.B.} \bsnm{{Whitmore}}},
\bauthor{\binits{M.T.} \bsnm{{Murphy}}},
\batitle{{Impact of instrumental systematic errors on fine-structure constant
  measurements with quasar spectra}}.
\bjtitle{\mnras}
\bvolume{447},
\bfpage{446}--\blpage{462}
(\byear{2015})
\end{barticle}
\endbibitem

\bibitem[\protect\citeauthoryear{{Whitmore} et~al.}{2010}]{Whitmore2010}
\begin{barticle}
\bauthor{\binits{J.B.} \bsnm{{Whitmore}}},
\bauthor{\binits{M.T.} \bsnm{{Murphy}}},
\bauthor{\binits{K.} \bsnm{{Griest}}},
\batitle{{Wavelength calibration of the VLT-UVES spectrograph}}.
\bjtitle{\apj}
\bvolume{723},
\bfpage{89}--\blpage{99}
(\byear{2010})
\end{barticle}
\endbibitem

\bibitem[\protect\citeauthoryear{Wilken et~al.}{{2012}}]{Wilken2012}
\begin{barticle}
\bauthor{\binits{T.} \bsnm{Wilken}},
\bauthor{\binits{G.} \bsnm{Lo~Curto}},
\bauthor{\binits{R.A.} \bsnm{Probst}},
\bauthor{\binits{T.} \bsnm{Steinmetz}},
\bauthor{\binits{A.} \bsnm{Manescau}},
\bauthor{\binits{L.} \bsnm{Pasquini}},
\bauthor{\binits{J.I.} \bsnm{Gonzalez~Hernandez}},
\bauthor{\binits{R.} \bsnm{Rebolo}},
\bauthor{\binits{T.W.} \bsnm{H\"{a}nsch}},
\bauthor{\binits{T.} \bsnm{Udem}},
\bauthor{\binits{R.} \bsnm{Holzwarth}},
\batitle{{A spectrograph for exoplanet observations calibrated at the
  centimetre-per-second level}}.
\bjtitle{Nature}
\bvolume{{485}},
\bfpage{611}--\blpage{614}
(\byear{{2012}})
\end{barticle}
\endbibitem

\bibitem[\protect\citeauthoryear{{Xu} et~al.}{2013}]{Xu2013}
\begin{barticle}
\bauthor{\binits{S.} \bsnm{{Xu}}},
\bauthor{\binits{M.} \bsnm{{Jura}}},
\bauthor{\binits{D.} \bsnm{{Koester}}},
\bauthor{\binits{B.} \bsnm{{Klein}}},
\bauthor{\binits{B.} \bsnm{{Zuckerman}}},
\batitle{{Discovery of molecular hydrogen in white dwarf atmospheres}}.
\bjtitle{\apjl}
\bvolume{766},
\bfpage{18}
(\byear{2013})
\end{barticle}
\endbibitem

\end{thebibliography}
\end{document}